\documentclass[reprint, twocolumn, amsfonts, amsmath, amssymb, superscriptaddress, aps, prl]{revtex4-2}

\usepackage{graphicx}
\usepackage{dcolumn}
\usepackage{bm}
\usepackage[normalem]{ulem}
\usepackage{mathtools}
\usepackage[dvipsnames]{xcolor}
\usepackage[linkcolor=Blue,citecolor=Blue,urlcolor=Blue,colorlinks=true]{hyperref}
\usepackage[normalem]{ulem}

\newif\ifshownotes
\ifshownotes
	\newcommand{\note}[3]{{\color{#2}[#1: #3]}}
    
    \newcommand{\del}[3]{\textbf{\color{#2}\sout{#3}}}
    \newcommand{\eqdel}[3]{\textbf{\color{#2}#3}}
\else
	\newcommand{\note}[3]{}
	\newcommand{\del}[3]{}
	\newcommand{\eqdel}[3]{}
    
\fi

\usepackage{xr}

\DeclareMathOperator{\tr}{tr}

\begin{document}
\title{Hidden nonreciprocity as a stabilizing effective potential in active matter}
\author{Matthew Du}
\email{madu@uchicago.edu}
\affiliation{Department of Chemistry, University of Chicago, Chicago, Illinois 60637, USA}
\affiliation{The James Franck Institute, University of Chicago, Chicago, Illinois 60637, USA}
\author{Andriy Goychuk}
\altaffiliation[Present address: ]{Department of Systems Immunology, Helmholtz Centre for Infection Research, 38124 Braunschweig, Germany}
\altaffiliation[Present address: ]{Lower Saxony Center for Artificial Intelligence and Causal Methods in Medicine (CAIMed), Hannover, Germany}
\altaffiliation[Present address: ]{Institute for Biochemistry, Biotechnology and Bioinformatics, Technische Universität Braunschweig, Braunschweig, Germany}
\affiliation{Institute for Medical Engineering and Science, Massachusetts Institute of Technology, Cambridge, Massachusetts 02139, USA}
\author{Suriyanarayanan Vaikuntanathan}
\email{svaikunt@uchicago.edu}
\affiliation{Department of Chemistry, University of Chicago, Chicago, Illinois 60637, USA}
\affiliation{The James Franck Institute, University of Chicago, Chicago, Illinois 60637, USA}
\date{\today}
\begin{abstract}
Nonreciprocal interactions are known to produce distinctive dynamics in active matter. 
To shed light on how the stationary state of such systems is affected by breaking reciprocity, we consider interacting particles propelled by persistent noise, where reciprocity is broken by a transverse 
force perpendicular to the gradient of the interaction energy.
Focusing on the steady-state distribution of positions, we show that the nonreciprocal coupling helps keep the system at its stable configurations.
Specifically, we demonstrate this effect for a variety of active systems whose stable configurations are energy minima, finding that the nonreciprocal coupling stiffens springs, aligns spins, and improves associative memory.
In contrast, the transverse force does not change the stationary distribution of positions at all when the noise is thermal. 
Preliminary simulations suggest that this nonreciprocal coupling plays a similar stabilizing role in other active systems, such as those exhibiting motility-induced phase separation, whose stable configurations are not energy minima.
\end{abstract}
\maketitle
\section{Introduction}
Active matter is a broad class of physical systems driven out of equilibrium by energy-consuming autonomous agents 
~\cite{ramaswamy_mechanics_2010, marchetti_hydrodynamics_2013, bechinger_active_2016, obyrne_time_2022}, leading to a diversity of single-particle and collective phenomena. 
It is known that the self-propulsion of active agents such as bacteria can lead to phase separation~\cite{cates_motility-induced_2015}. 
More recently, it was shown that nonreciprocal interactions, for example chemophoretic couplings which violate Netwon's third law in nonidentical colloidal particles~\cite{soto_self-assembly_2014, agudo-canalejo_active_2019}, can lead to distinctive dynamics, including oscillatory collective motion~\cite{fruchart_non-reciprocal_2021, you_nonreciprocity_2020, saha_scalar_2020, tan_odd_2022, frohoff_2021, brauns_2024} and unidirectional energy propagation~\cite{brandenbourger_non-reciprocal_2019, saha_scalar_2020}.  

The effect of nonreciprocity on the stationary state of active systems is less clear.
A possible outcome is that nonreciprocal interactions induce divergence-free phase space flows which, similar to the chiral diffusion of non-interacting particles~\cite{hargus_odd_diffusion_2021}, do not affect the macroscopic distribution of particles.
Yet, there are also examples where
breaking the reciprocity of interactions can alter the steady-state distribution of particle positions~\cite{bartnick_structural_2015, chiu_phase_2023, sinha_how_2024, martin_exact_2023, dinelli_non-reciprocity_2023, cure_antagonistic_2023, loos_long-range_2023} and velocities~\cite{ivlev_statistical_2015}.
Furthermore, the stationary state could also be influenced by an interplay between nonreciprocal interactions and self-propulsion~
\cite{duan_dynamical_2023, martin_exact_2023, dinelli_non-reciprocity_2023}.

\begin{figure}
\centering\includegraphics{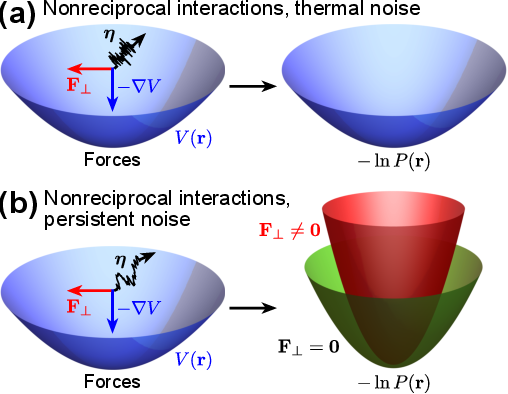}
\caption{Illustration of the steady-state distribution $P(\mathbf{r})$ of active Ornstein-Uhlenbeck particles. 
The persistence in the noise $\bm\eta$ (black jagged arrows) captures the propulsion of the particles. 
The potential force $-\nabla V$ (blue arrow; the blue paraboloid shows the potential $V$) models reciprocal interactions, while the transverse force $\mathbf{F}_\perp \perp -\nabla V$ (red arrow) models nonreciprocal interactions.
(a) In the limit of vanishing persistence, the noise is thermal, and $P(\mathbf{r})$ is
the Boltzmann distribution determined by $V$ and independently of $\mathbf{F}_\perp$.
(b) With persistence in the noise,  $\mathbf{F}_\perp$ appears in $P(\mathbf{r})$ as an effective energy correction that helps keep the system at its stable configurations.
\label{fig:schematic}}
\end{figure}

To shed light on these phenomena, we study nonreciprocally interacting active particles where the reciprocity is broken by a transverse force, which is perpendicular to the potential gradient representing the reciprocal part of the interactions. 
Analytical calculations, together with simulations, paint a picture where nonreciprocity stabilizes self-propelling matter around its most likely configurations.
For a variety of examples, we find an effective steepening of the energy basins containing potential minima [Fig. \ref{fig:schematic}(b)].
We also present simulations suggesting that a similar stabilization might occur in systems whose stable configurations are not energy minima, e.g., particles undergoing motility-induced phase separation (MIPS)~\cite{cates_motility-induced_2015}.
The stabilizing effect disappears in the limit of vanishing persistence time~\cite{zhou_quasi-potential_2012} [Fig. \ref{fig:schematic}(a)].
These key results have direct implications for the inference of nonreciprocal interactions in active matter~\cite{chen_generalized_2023, chardes_stochastic_2023, yu_physics_2025} and the engineering of such couplings to produce properties of interest~\cite{brandenbourger_non-reciprocal_2019, osat_non-reciprocal_2022}.    

\section{Model}
We demonstrate our results in the context of a prototypical model of active matter, namely, active Ornstein-Uhlenbeck particles (AOUPs)~\cite{sepulveda_collective_2013, szamel_self-propelled_2014, koumakis_directed_2014, bonilla_active_2019, martin_statistical_2021}. 
For a system of $N$ AOUPs, the collective spatial coordinates $\mathbf{r}= (\mathbf{r}_1,\dots,\mathbf{r}_N)$ follow the overdamped Langevin equation
\begin{equation}
\dot{\mathbf{r}} = 
\mathbf{F}(\mathbf{r}) + \bm \eta (t) \,.
\label{eq:eom}
\end{equation}
The persistent Gaussian noise $\bm{\eta}$ has zero mean, $\langle \eta_{i} (t)\rangle = 0$, and covariance $\langle \eta_{i} (t) \eta_{j} (t') \rangle 
= \frac{T}{\tau} \delta_{ij} e^{-|t-t'|/\tau}$ featuring the persistence time $\tau$ and noise strength $T$ to represent particle self-propulsion and fluctuations of the bath.
We decompose the particle-particle interactions,
\begin{equation}
\mathbf{F} = -\nabla V + \mathbf{F}_\perp,
\label{eq:f-norm-decomp}
\end{equation}
into the conservative force $-\nabla V$ associated with the potential $V(\mathbf{r})$ and the non-conservative force $\mathbf{F}_\perp(\mathbf{r})$ that breaks reciprocity.
Following Ref.~\cite{zhou_quasi-potential_2012}, we consider the normal decomposition
\begin{align}
-\nabla V \cdot \mathbf{F}_{\perp} &= 0,
\label{eq:perp}\\
\nabla \cdot \mathbf{F}_\perp &= 0,
\label{eq:div-free}
\end{align}
so that $\mathbf{F}_\perp$ is transverse to $-\nabla V$ and has zero divergence, respectively. 
This partitioning has the physical interpretation that the quasi-potential $V$ is a good measure of global stability but may not always be possible for general forces~\cite{zhou_quasi-potential_2012}.
Nevertheless, we will discuss below a simple way to construct, for any $V$, a $\mathbf{F}_\perp$ satisfying Eqs. \eqref{eq:perp}-\eqref{eq:div-free}. 
Nonreciprocity is characterized here by
\begin{equation}
\frac{\partial F_{\perp,i}}{\partial r_j} 
\neq \frac{\partial F_{\perp,j}}{\partial r_i},\quad i\neq j,
\label{eq:nr}
\end{equation}
where $i,j$ run over all particles and their respective spatial coordinates.
As a control, we also consider the limit $\tau \rightarrow 0$, where $\bm \eta$  reduces to thermal noise, which has zero mean and no temporal correlations, $\langle \eta_{i} (t) \eta_{j} (t') \rangle = 2T \delta_{ij} \delta(t-t')$.
For any $\mathbf{F}$ of the form \eqref{eq:f-norm-decomp}--\eqref{eq:div-free}, it is known that the steady-state distribution of positions under thermal noise is the Boltzmann distribution $P(\mathbf{r}) \propto \exp(-V/T)$~\cite{zhou_quasi-potential_2012}.
Thus, nonreciprocity in the form of $\mathbf{F}_\perp$ has no effect on the distribution in the thermal limit [Fig. \ref{fig:schematic}(a)]. 
To intuitively understand why, we consider the dynamics if driven by $\mathbf{F}_\perp$ alone, $\dot{\mathbf{r}} = \mathbf{F}_{\perp}$.
Using its transverse nature [Eq. \eqref{eq:perp}] and the chain rule, we find that the potential energy of the system is conserved, $\dot{V} = 0$.
Since $\mathbf{F}_{\perp}$ has zero divergence [Eq. \eqref{eq:div-free}], then trajectories in general do not terminate (at a fixed point) \cite{strogatz_nonlinear_2018}.
Hence, $\mathbf{F}_\perp$ on its own leads to indefinite switching among configurations that have equal potential energy, which all have the same thermal probability, providing intuition for why this nonreciprocal force does not alter $P(\mathbf{r})$ under thermal noise.

Throughout the remainder of this work, we focus on specific systems of AOUPs, each defined by a different choice of interparticle force $\mathbf{F}$ [Eq.~\eqref{eq:eom}], in which the nonreciprocity is controlled by a single parameter $\alpha$. 
That is, $\alpha \neq 0$ breaks reciprocity ($\mathbf{F}_\perp \neq \mathbf{0}$) while $\alpha=0$ does not ($\mathbf{F}_\perp = \mathbf{0}$).
Details related to the simulation of each system can be found in Appendix~\ref{app:numerical}.

\section{Nonreciprocal harmonic oscillators} 
\label{sec:harmonic}
\begin{figure}
\centering\includegraphics{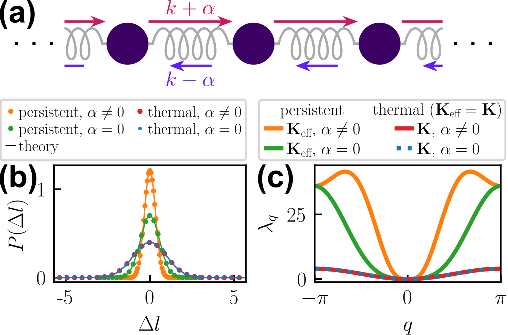}
\caption{%
Nonreciprocal spring-mass chain with periodic boundary conditions. (a) Schematic diagram.
For each pair of neighboring masses, the left (right) mass acts on the right (left) mass with force constant $k+\alpha$ ($k-\alpha$). 
(b) Steady-state distribution $P(\Delta l)$ of spring displacements $\Delta l$ for various noise types and $\alpha=0,2$. 
Simulations (circles) match theoretical predictions (lines) for different conditions (color code).
(c) Eigenvalues $\lambda_q$, for each wavenumber $q$, of the effective force constant matrix $\mathbf{K}_\text{eff}$ governing $P(\mathbf{r})$ for various noise types and values of $\alpha=0,2$.
In (b) and (c), the other parameters are $N=100$, $k = 1$, $T=1$, and $\tau = 2$. 
\label{fig:harmonic}}
\end{figure}

We begin with the simple example of a periodic spring-mass chain where the springs nonreciprocally couple neighboring masses [Fig. \ref{fig:harmonic}(a)].
Previously, this system has been studied in the underdamped limit~\cite{brandenbourger_non-reciprocal_2019}. 

Mass $i$ acts on mass $i+1$ with spring constant $k+\alpha$ and reacts with a \emph{different} spring constant $k-\alpha$, thereby violating Newton's third law.
The force experienced by mass $i=1,\dots,N$ is
\begin{equation}
F_i = -\sum_{s=\pm 1}(k-s\alpha)(r_i - r_{i+s}),
\label{eq:f-lin-1}
\end{equation}
where $r_i$ is the displacement of mass $i$ from its rest position, and periodic boundary conditions are imposed as $r_{N+1} \equiv r_1$ and $r_0 \equiv r_N$.
The force $\mathbf{F}$ satisfies the normal decomposition [Eqs.~\eqref{eq:f-norm-decomp}-\eqref{eq:div-free}] 
\begin{align}
    V&=\sum_{i=1}^{N}\frac{1}{2}k(r_i - r_{i+1})^2, \label{eq:v-lin-1}\\
    F_{\perp,i} &= \alpha \sum_{s=\pm 1}s(r_i - r_{i+s}),\label{eq:f-perp-lin-1}
\end{align}
which shows that the nonreciprocity arises from the transverse force $\mathbf{F}_\perp$.
In the continuum limit and multiple spatial dimensions, the nonreciprocal system converges to an active ring polymer whose monomers self-propel tangentially to the backbone with force $\propto \alpha$~\cite{bianco_globulelike_2018,winkler_2020}.
Specifically, in dimensions $d\geq 2$, the tensile force due to the curvature of the polymer is perpendicular to the tangent along its backbone.
Hence, the active force is transverse to the potential force at every point in the body of the chain.
This transversality is broken in the case of open chains, because the potential forces and the tangential forces are parallel at the chain tips, and by interactions between non-neighboring monomers.

To gain intuition for the nonreciprocal coupling, we consider the dynamics driven by only the nonreciprocal transverse force,  $\dot{\mathbf{r}} = \mathbf{F}_\perp$.
We analytically solve for the state at time $t$,
\begin{align}
    \mathbf{r}(t) = 
    &\; r_0(0) \mathbf{v}_0 
    + 
    \begin{cases}
        0, & N\text{ odd} \\
        r_{\pi}(0) \mathbf{v}_{\pi}, & N\text{ even}
    \end{cases}\nonumber\\     
    &+\sum_{q\neq 0, \pi} r_q(0) \left\{\cos\left[2\alpha(\sin q)t\right]\mathbf{v}_q\right.\nonumber\\
    &\hspace{6em}\left.+ \sin\left[2\alpha(\sin q)t\right]\mathbf{v}_{-q} \right\}
    \label{eq:revec-lin-1}
\end{align}
where 
\begin{equation}
    \mathbf{v}_{q}=
    \begin{cases}
        \mathbf{u}_{q}, & q=0,\pi\\
        \frac{1}{\sqrt{2}}\left(\mathbf{u}_{q}-\mathbf{u}_{-q}\right), & 0<q<\pi\\
        \frac{1}{i\sqrt{2}}\left(\mathbf{u}_{|q|}-\mathbf{v}_{-|q|}\right), & -\pi<q<0
    \end{cases}
\end{equation}
are real-valued normal modes of the reciprocal system ($\dot{\mathbf{r}} = -\nabla V$), $\mathbf{r}(0)=\sum_q r_q(0) \mathbf{v}_q$ is the initial state in this basis, $u_{qj} = e^{iqj}/\sqrt{N}$ are complex-valued normal modes of the reciprocal system, and the wavenumbers $q \in (-\pi, \pi]$ take the $N$ values for which $e^{iqN}$ is an $N$th root of unity. 
We see that the nonreciprocal coupling forces the system to oscillate at frequency $\propto |\alpha|$ between degenerate normal modes $\mathbf{v}_{\pm q}$ of the reciprocal system.

Fig.~\ref{fig:harmonic}(b) shows the stationary state as determined from simulations. 
As expected from the normal decomposition of the interparticle force, under thermal noise, nonreciprocal and reciprocal springs have the same steady-state distribution of displacements from rest length. 
Under persistent noise, however, the nonreciprocal springs have smaller displacements than the reciprocal springs.
Thus, persistent noise enables nonreciprocity to stabilize the springs at their minimal-energy configurations.

To understand this stabilizing effect, we consider a general harmonic oscillator, where $\mathbf{F} = -\mathbf{C}\mathbf{r}$ is the interparticle force,
$\mathbf{C}$ is the force constant matrix, $\mathbf{r}$ indicates the displacement of each particle from its rest position, and the nonreciprocity~\eqref{eq:nr} corresponds to an asymmetry $\mathbf{C}\neq \mathbf{C}^T$.
The normal decomposition \eqref{eq:f-norm-decomp}--\eqref{eq:div-free} exists and is given by~\cite{kwon_structure_2005, ao_stochastic_2003, noh_steady-state_2015} $V = \frac{1}{2}\mathbf{r}^T \mathbf{K} \mathbf{r}$ and $\mathbf{F}_\perp = -\mathbf{A}\mathbf{r}$,
where $\mathbf{K}$ is symmetric, $\mathbf{A}$ is traceless, and $\mathbf{K}\mathbf{A}$ is antisymmetric.
Following standard approaches of solving linear stochastic differential equations~\cite{gardiner_handbook_1985,fodor_how_2016}, we derive (see Appendix~\ref{app:prob-harmonic}) the exact steady-state distribution of positions under persistent noise, $P(\mathbf{r}) \propto \exp(-V_\text{eff}/T)$, where $V_\text{eff}(\mathbf{r}) = \frac{1}{2}\mathbf{r}^T \mathbf{K}_\text{eff} \mathbf{r}$ is a harmonic potential with effective force constant matrix
\begin{equation}
\mathbf{K}_\text{eff} = 
\mathbf{K}+\tau\mathbf{K}^T \mathbf{K}
+\tau\mathbf{A}^{T}\left[\mathbf{I}-\left(\mathbf{I}+\tau\mathbf{K}\right)^{-1}\right]\mathbf{A}.
\label{eq:k-eff}
\end{equation}
The first term of $\mathbf{K}_\text{eff}$ corresponds to the potential $V$, whose minima satisfy $V(\mathbf{r})=0$.
These minima are stabilized by persistent noise as shown in the second term~\cite{maggi_multidimensional_2015, fodor_how_2016,bonilla_active_2019}, independently of the transverse force.
The third term arises from a synergy between this nonreciprocal coupling and persistent noise, and scales as $\mathcal{O}(\tau^2)$.   

We will now prove that the third, nonreciprocal term of $\mathbf{K}_\text{eff}$ [Eq.~\eqref{eq:k-eff}] provides additional stabilization to the minima of $V$. 
Summarized below, we present the full proof in Appendix~\ref{app:third-term-stabilizes}.
Considering that the system is stable under thermal noise, $\mathbf{K}$ is positive semidefinite.
Hence, the nonreciprocal term is also positive semidefinite and, if we assume the generally true condition $\mathbf{K}\mathbf{A}\neq \mathbf{0}$, nonzero.
So, this term increases at least one eigenvalue of the effective force constant matrix $\mathbf{K}_\text{eff}$ while leaving the remaining eigenvalues unchanged.
Using this result (along with others), we can also show that the nonreciprocal term preserves the set of zero modes of $\mathbf{K}_\text{eff}$, which are exactly the zero modes of $\mathbf{K}$. 
Thus, the transverse force that nonreciprocally couples the particles increases the likelihood of finding the system at the minima of $V$ [Fig.~\ref{fig:schematic}(b)].

We can make more precise statements about how the eigenvalues of $\mathbf{K}_\text{eff}$ [Eq.~\eqref{eq:k-eff}] are affected by its nonreciprocal term [third term of Eq.~\eqref{eq:k-eff}] if the force constant matrix $\mathbf{C}$ is normal. 
In this special case, $\mathbf{K}$ and $\mathbf{A}$ are the symmetric and antisymmetric parts of $\mathbf{C}$ \cite{noh_steady-state_2015}, respectively, and therefore they satisfy $[\mathbf{K},\mathbf{A}]=\mathbf{0}$ due to $\mathbf{C}$ being normal.
This commutation relation implies that $\mathbf{K}$, $\mathbf{A}$, and thus $\mathbf{K}_\text{eff}$ [see Eq. \eqref{eq:k-eff}] are simultaneously diagonalizable, allowing the eigenvalues of $\mathbf{K}_\text{eff}$ to be expressed in terms of those of $\mathbf{K}$ and $\mathbf{A}$,
\begin{align}
\lambda_{n}(\mathbf{K}_{\text{eff}})=&\;\lambda_{n}(\mathbf{K})+\tau\left[\lambda_{n}(\mathbf{K})\right]^{2}\nonumber\\
&+\tau\left|\lambda_{n}(\mathbf{A})\right|^{2}\left\{ 1-\left[1+\tau\lambda_{n}(\mathbf{K})\right]^{-1}\right\},
\label{eq:eval-k-eff}
\end{align}
where $\lambda_n (\mathbf{M})$ denotes the eigenvalue of $\mathbf{M} \in \{ \mathbf{K},\mathbf{A},\mathbf{K}_\text{eff} \}$ associated with the $n$th simultaneous eigenvector, and the index $n$ can be based on any ordering of the eigenvectors.
The third term of Eq. \eqref{eq:eval-k-eff} is the contribution of the nonreciprocal term to the eigenvalues of $\mathbf{K}_{\text{eff}}$.
We see that the nonreciprocal term increases the $n$th eigenvalue of $\mathbf{K}_{\text{eff}}$ if the corresponding eigenvalues of $\mathbf{K}$ and $\mathbf{A}$ are both nonzero.
Otherwise, the nonreciprocal term has no effect on $\lambda_{n}(\mathbf{K}_{\text{eff}})$.
In Appendix~\ref{app:avg-pot}, we additionally demonstrate, by studying the dynamics of the noise-averaged potential energy for normal force constant matrices $\mathbf{C}$, that the nonreciprocal force concomitantly reduces the energy injected into the system by persistent noise.

We will now apply the above theory for general harmonic oscillators to analytically calculate the steady-state properties of the nonreciprocal spring-mass chain [Fig. \ref{fig:harmonic}(a)].
Expressing the total force [Eq. \eqref{eq:f-lin-1}] in the form $\mathbf{F}=-\mathbf{C}\mathbf{r}$ and noticing that $\mathbf{C}$ is a normal matrix, we straightforwardly determine the matrices that satisfy the normal decomposition as $\mathbf{K} = (\mathbf{C} + \mathbf{C}^T)/2$ and $\mathbf{A}= (\mathbf{C} - \mathbf{C}^T)/2$.
Using Eq. \eqref{eq:k-eff} for the effective force constant matrix $\mathbf{K}_\text{eff}$ that governs $P(\mathbf{r})$ under persistent noise, we calculate the steady-state distribution of spring displacements to have zero mean and a variance of
\begin{widetext}
\begin{equation}
\text{Var}(\Delta l)
= \frac{T}{N}\sum_{q\neq0}\frac{1}{k+2\tau k^{2}(1-\cos q)+2\tau \alpha^{2}(1+\cos q)\left\{ 1-\left[1+2\tau k(1-\cos q)\right]^{-1}\right\} }.
\label{eq:var-displ-lin-1}
\end{equation}
\end{widetext}
As shown in Fig. \ref{fig:harmonic}(b), this analytical distribution reproduces exactly the simulated distribution under persistent noise ($\tau\neq 0$) for different values of nonreciprocity $\alpha$, and it reduces to the correct distribution under thermal noise ($\tau\rightarrow 0$).
Continuing the theoretical analysis, we calculate the eigenvalues of $\mathbf{K}_\text{eff}$ using Eq.~\eqref{eq:eval-k-eff} for normal force constant matrix $\mathbf{C}$, 
\begin{align}
\lambda_q (\mathbf{K}_\text{eff}) = 
&\;2k(1-\cos q)
+4\tau k^{2}(1-\cos q)^{2}\nonumber\\
&+4\tau \alpha^{2}\sin^{2}q\left\{ 1-\left[1+2\tau k(1-\cos q)\right]^{-1}\right\},
\label{eq:eval-lin-1}
\end{align} 
which are plotted in Fig. \ref{fig:harmonic}(c).
As shown above for general harmonic oscillators, at most wavenumbers, persistent noise raises the associated eigenvalue for the reciprocal springs and even more for the nonreciprocal springs,
while, at the other wavenumbers, the eigenvalues are unchanged. 

To highlight the generality of our results, we show in Appendix~\ref{app:robust} that the nonreciprocity-driven stabilization of the one-dimensional nonreciprocal spring-mass chain [Fig.~\ref{fig:harmonic}(b)] is robust to disorder and open boundary conditions.
Also, in Appendix~\ref{app:dusty-plasma}, we further validate the general harmonic-oscillator theory, i.e., Eq.~\eqref{eq:k-eff}, using a spring-mass model of dusty plasma \cite{melzer_structure_1996}, where reciprocity is broken by a force not transverse to the interaction potential gradient. 

\section{Nonreciprocal spherical models}
\label{sec:spherical}
\begin{figure}
\centering\includegraphics{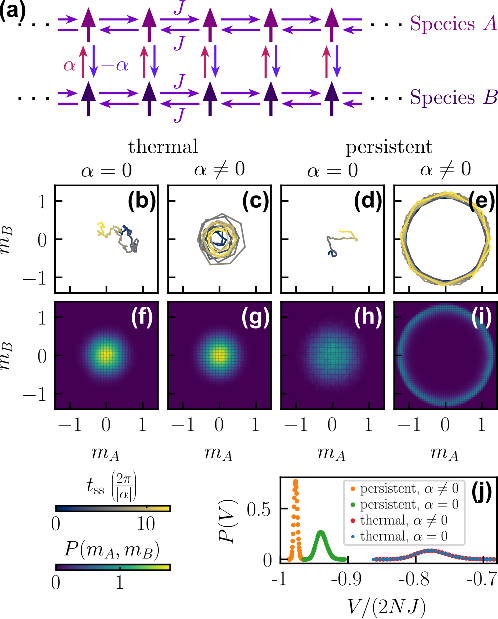}
\caption{Nonreciprocal spherical model with two species of spins. 
(a) Schematic diagram.
Spins of species $A$ and $B$ are reciprocally coupled to neighboring spins of the same species with strength $J$.
In contrast, $A$ spins align with $B$ spins with strength $\alpha$, whereas $B$ spins (anti-)align with $A$ spins with strength $-\alpha$. 
(b-g) Steady-state dynamics (b-e) and distribution $P(m_A,m_B)$ (f-i) of the magnetizations $m_A$ and $m_B$ of $A$ and $B$ spins, respectively, under thermal (b, c, f, g) and persistent (d, e, h, i) noise and for $\alpha = 0$ (b, d, f, h) and $\alpha = 2$ (c, e, g, i).
In (b-e), the results shown are from the final 100 time points $t_\text{ss}$ at which the spin values are stored during the long-time simulations used to compute steady-state properties (Appendix~\ref{app:numerical}).
(j) Steady-state distribution of potential energy, $P(V)$, for various noise types and values of $\alpha=0,2$.
In (b)-(j), the other parameters are $N=100$, $J = 1$, $\mu = 2$, $T=0.5$, and $\tau = 2$. 
\label{fig:bipartite-spherical}}
\end{figure}

Next, we demonstrate the stabilizing nature of transverse forces for nonlinear $\mathbf{F}$. 
We consider a nonreciprocal version of the spherical model~\cite{berlin_spherical_1952} with continuous spins of species $A$ and $B$ [Fig.~\ref{fig:bipartite-spherical}(a)].
This extends the discrete Ising model of Ref.~\cite{avni_non-reciprocal_2023} and can be readily adapted to represent polar alignment in flocks~\cite{tonertu_1995, loos_long-range_2023, solon_revisiting_2013, solon_flocking_2015}.

Spin $i=1,\dots,N$ of species $S\in\{A,B\}$ feels a force
\begin{subequations}
\begin{align}
F_{A,i} 
&= -\mu \left(|\bm\sigma|^2- 2N\right)\sigma_{A,i}
+J\sum_{s=\pm 1}\sigma_{A,i+s} +\alpha \sigma_{B,i},\\
F_{B,i} 
&= -\mu \left(|\bm\sigma|^2- 2N\right)\sigma_{B,i}
+J\sum_{s=\pm 1}\sigma_{B,i+s} -\alpha \sigma_{A,i},
\end{align}
\label{eq:f-sph-2}%
\end{subequations}
respectively, where $\sigma_{S,i}$ is the value of the $i$th spin of species $S$, periodic boundary conditions are defined by $\sigma_{S,N+1} \equiv \sigma_{S,1}$ and $\sigma_{S,0} \equiv \sigma_{S,N}$,  $\bm \sigma$ is the vector of all spins, and $\mu$  is the hardness of the soft constraint that imposes the condition $|\bm \sigma|^2 = 2N$.
Within species, the nearest neighbors are coupled reciprocally with strength $J$.
Between species, the interaction is fully nonreciprocal with magnitude $|\alpha|$.
If $\alpha > 0$, for example, each $A$ spin aligns with its corresponding $B$ spin, whereas the $B$ spin antialigns with the $A$ spin.
The normal decomposition [Eq. \eqref{eq:f-norm-decomp}-\eqref{eq:div-free}] of $\mathbf{F}$ in Eq. \eqref{eq:f-sph-2} is given by
\begin{align}
V &= \frac{1}{4}\mu \left(|\bm\sigma|^2- 2N\right)^2 -\sum_{S,i} J\sigma_{S,i}\sigma_{S,i+1}, \label{eq:v-sph-2}\\
F_{\perp,S,i} &= \alpha \times
\begin{cases}
\sigma_{B,i}, & S=A,\\
-\sigma_{A,i}, & S=B,
\end{cases}\label{eq:f-perp-sph-2}
\end{align}
revealing that the nonreciprocal coupling can be associated with the transverse force.  

The nonreciprocal coupling [Eq.~\eqref{eq:f-perp-sph-2}] induces time-periodic swapping of magnetization between the two species, as in the nonreciprocal Ising model~\cite{avni_non-reciprocal_2023}.
We can understand this effect analytically by considering the dynamics of the system driven only by the reciprocity-breaking, transverse force, $\dot{\bm{\sigma}} = \mathbf{F}_\perp$.
Consider the initial state $\bm\sigma (0) = \big(\sigma_{A,1}(0),\dots,\sigma_{A,N}(0),\sigma_{B,1}(0),\dots,\sigma_{B,N}(0)\big) \equiv \big(\bm\sigma_A(0), \bm\sigma_B(0)\big)$.
Taking advantage of the linearity of $\mathbf{F}_\perp$ [Eq.~\eqref{eq:f-perp-sph-2}],
we obtain the state at time $t$ as
\begin{equation}
\bm{\sigma}(t) = 
\begin{pmatrix}\cos\alpha t & \sin\alpha t\\
-\sin\alpha t & \cos\alpha t
\end{pmatrix}\begin{pmatrix}\bm\sigma_{A}(0)\\
\bm\sigma_{B}(0)
\end{pmatrix}.
\label{eq:periodic-spins}
\end{equation}
We see that the nonreciprocal coupling drives the two species of spins to periodically swap configurations at frequency $\propto|\alpha|$, in a way that conserves the potential energy $V$ [Eq.~\eqref{eq:v-sph-2}].  
This oscillatory dynamics manifests as a periodic swapping of magnetization $m_S=N^{-1}\sum_{i=1}^N \sigma_{S,i}$ between species $S=A,B$,   
\begin{equation}
\begin{pmatrix}m_{A}(t)\\
m_{B}(t)
\end{pmatrix}=\begin{pmatrix}\cos\alpha t & \sin\alpha t\\
-\sin\alpha t & \cos\alpha t
\end{pmatrix}\begin{pmatrix}m_{A}(0)\\
m_{B}(0)
\end{pmatrix}.
\end{equation}

Returning to the system with reciprocal coupling and noise, we find that the nonreciprocal coupling also produces swapping of magnetization between $A$ and $B$ spins under both thermal [Fig.~\ref{fig:bipartite-spherical}, cf. (b) and (c)] and persistent noise [Fig.~\ref{fig:bipartite-spherical}, cf. (d) and (e)]. 
In contrast, nonreciprocity has no effect on the overall magnetization of all spins, $|(m_A,m_B)|$, under thermal noise [Fig.~\ref{fig:bipartite-spherical}, cf. (b) and (c)], but it causes $|(m_A,m_B)|$ to have a higher value and less fluctuations under persistent noise [Fig.~\ref{fig:bipartite-spherical}, cf. (d) and (e)].
In the language of the nonreciprocal Ising model~\cite{avni_non-reciprocal_2023}, our nonreciprocal system under persistent noise [Fig.~\ref{fig:bipartite-spherical}(e)] but not thermal noise [Fig.~\ref{fig:bipartite-spherical}(c)] is in the `swap phase', where $(m_A, m_B)$ rotates about $(0, 0)$ and maintains a norm $\not\approx 0$. 
While it would be interesting to see whether the `swap phase' survives for larger system size $N$, we defer this question, which has been explored for the nonreciprocal Ising model~\cite{avni_non-reciprocal_2023}, to future studies. 

How nonreciprocity affects the stationary state reflects its effect on the dynamics.
Indeed, we find that nonreciprocity does not modify the stationary properties under thermal noise [Fig.~\ref{fig:bipartite-spherical}, cf. (f) and (g)], as expected from the normal decomposition of $\mathbf {F}$, whereas, under persistent noise, nonreciprocity simultaneously increases the magnetization of both species [Fig.~\ref{fig:bipartite-spherical}, cf. (h) and (i)] and lowers the total potential energy [Fig.~\ref{fig:bipartite-spherical}(j)].
Therefore, persistent noise enables nonreciprocity to stabilize the maximally aligned configurations, i.e., global energy minima.
For completeness, we show in Appendix~\ref{app:spherical-1} that this principle is corroborated for spins of a single species with nonreciprocal interactions, even in presence of a magnetic field or antiferromagnetic coupling.

\section{A general construction of transverse nonreciprocal forces}
Thus far, we have focused on essentially one-dimensional systems where simple modifications to local interactions can produce a reciprocity-violating transverse force.
However, it is unclear if this outcome extends to arbitrarily complex systems. 
To apply our findings more broadly, we consider a transverse force of the form 
\begin{equation}
\mathbf{F}_\perp = \mathbf{A}(-\nabla V),
\label{eq:antisym}
\end{equation}
where $\mathbf{A}$ is an antisymmetric matrix.
Eq.~\eqref{eq:antisym} automatically satisfies properties \eqref{eq:perp}-\eqref{eq:nr} of a transverse force that nonreciprocally couples different degrees of freedom.
Similar approaches have been used for faster sampling of steady-state distributions~\cite{ghimenti_sampling_2023}.
If one knows $-\nabla V$, then an antisymmetric linear transformation provides a straightforward (in principle) and general way to construct $\mathbf{F}_\perp$.

\section{Nonreciprocal spherical Hopfield model}
\label{sec:hopfield}
\begin{figure}
\centering\includegraphics{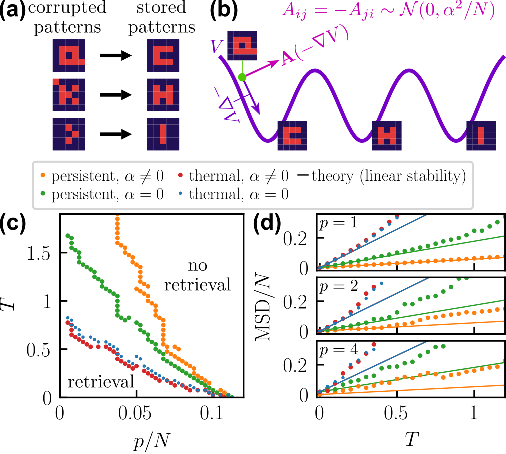}
\caption{%
Nonreciprocal spherical Hopfield model of associative memory.  
(a, b) Schematic diagram.
(a) Associative memory. 
Stored patterns can be retrieved from corrupted versions. 
(b) In the Hopfield model, the coupling among $N$ spins encodes $p$ random patterns.
For small pattern loading $p/N$, the patterns are stored as minima of the potential $V$ (purple curve) and can be retrieved by gradient descent (purple arrow).
We introduce nonreciprocity via a transverse force, $\mathbf{A}(-\nabla V)$ (magenta arrow), with $A_{ij}=-A_{ji} \sim \mathcal{N}(0, \alpha^2 / N)$ a random antisymmetric matrix.
(c, d) We compare various quantities for thermal versus persistent noise and different values of $\alpha=0,4$.
(c) Phase diagram of pattern retrieval in simulations, as a function of $T$ and $p/N$. 
Points of the same color form a transition line, below which the stored patterns can be retrieved and above which they cannot. 
(d) MSD from the energy minimum corresponding to the target pattern.
Circles show values from retrieval simulations after an initial relaxation of the system, while lines are calculated from a linear stability analysis based on Eq.~\eqref{eq:k-eff}. 
The other parameters are $N = 200$, $\lambda = 2$, $u_0 = 1$, and $\tau = 1$.
\label{fig:hopfield}}
\end{figure}

To assess whether breaking reciprocity with transverse force [Eq.~\eqref{eq:antisym}] can enhance stability, we first apply this nonreciprocal coupling to a spherical \cite{bolle_spherical_2003} Hopfield model of associative memory \cite{hopfield_neural_1982} [Fig.~\ref{fig:hopfield}, (a)-(b)].
In this biologically inspired disordered network, information is retrieved via convergence toward stable configurations---stored patterns realized as energy minima---making it a natural testbed for exploring stabilization through nonreciprocity.
Previously, the Hopfield model has been studied with nonreciprocal interactions~\cite{hopfield_neural_1982, hertz_memory_1986, parisi_asymmetric_1986, sompolinsky_temporal_1986, derrida_exactly_1987, crisanti_dynamics_1987, feigelman_augmented_1987, treves_metastable_1988, singh_fixed_1995, osat_non-reciprocal_2022} and persistent noise~\cite{behera_enhanced_2023, du_active_2024} separately but not together.

A fully connected network of $N$ spins encode $p$ random patterns through their disordered (reciprocal) couplings, which we describe with the potential of Ref.~\cite{bolle_spherical_2003} except that we use a soft spherical constraint: 
\begin{align}
V = 
&\;\frac{\lambda}{4}\left(|\bm{\sigma}|^{2}-N\right)^{2}\nonumber\\
&-\frac{1}{2}\sum_{ij}J_{ij}\sigma_{i}\sigma_{j}
-\frac{u_0}{4}\sum_{ijkl}J_{ijkl}\sigma_{i}\sigma_{j}\sigma_{k}\sigma_{l}.
\label{eq:v-hopfield}
\end{align}
The first term softly constrains the continuous spin vector $\bm{\sigma}=(\sigma_1,\dots,\sigma_N)$ to the sphere $|\bm\sigma|^2 = N$.
The second and third terms are the couplings that encode $p$ patterns $\{\bm{\xi}^{\mu}\}_{\mu=1}^{p}$ as (approximate) potential minima, where the pattern elements $\xi^{\mu}_i \sim \mathcal{N}(0, 1)$ are independent and identically distributed random variables.
In analogy to the standard Hopfield model \cite{hopfield_neural_1982}, the second term describes two-body interactions whose strengths are given by the connectivity matrix
\begin{equation}
J_{ij}=\begin{cases}
\frac{1}{N}\sum_{\mu=1}^{p}\xi_{i}^{\mu}\xi_{j}^{\mu}, & i\neq j\\
0, & i=j.
\end{cases}
\end{equation}
The third term---introduced by Ref. \cite{bolle_spherical_2003} to address the issue that the second term is not enough to stabilize continuous spins at the pattern configurations---describes 4-body interactions with the connectivity tensor
\begin{equation}
J_{ijkl}=\frac{1}{N^{3}}\sum_{\mu=1}^{p}\xi_{i}^{\mu}\xi_{j}^{\mu}\xi_{k}^{\mu}\xi_{l}^{\mu}.
\end{equation}
If the pattern loading $p/N$ is small enough~\cite{hopfield_neural_1982, bolle_spherical_2003}, then the patterns will be stored as minima of the interaction potential $V$ [Fig.~\ref{fig:hopfield}(b)]. 
A stored pattern can then be retrieved from a corrupted version of it [Fig.~\ref{fig:hopfield}(a)] by gradient descent [Fig.~\ref{fig:hopfield}(b)]. 

For the transverse force [Eq.~\eqref{eq:antisym}] that nonreciprocally couples the spins  [Fig.~\ref{fig:hopfield}(b)], we randomly choose the off-diagonal elements of $\mathbf{A}$ as $A_{ij} \sim \mathcal{N}(0, \alpha^2 / N)$~\cite{crisanti_dynamics_1987, singh_fixed_1995}.
As shown in Fig.~\ref{fig:hopfield}(c) and expected from our above findings, the transverse force does not significantly affect the ability to retrieve patterns under thermal noise, but it enables retrieval at higher noise strengths under persistent noise. 

To shed light on the enhanced retrieval, we carry out a linear stability analysis.
We first linearize the force about the energy minimum $\bm\sigma=\bm\sigma_{\min}$ corresponding to the target pattern: $\mathbf{F} \approx -\mathbf{C} (\bm{\sigma}-\bm{\sigma}_{\min})$.
We then calculate the steady-state distribution of spin values under persistent noise using the corresponding quantity presented above for nonreciprocal harmonic oscillators (Sec.~\ref{sec:harmonic}): $P(\bm{\sigma})\propto \exp\left[-(\bm{\sigma}-\bm{\sigma}_{\min})^T \mathbf{K}_\text{eff} (\bm{\sigma}-\bm{\sigma}_{\min})/(2T)\right]$, where $\mathbf{K}_\text{eff}$ is given by Eq.~\eqref{eq:k-eff}.
We compute $\mathbf{K}_\text{eff}$ from the $\mathbf{K}$ and $\mathbf{A}$ obtained by numerical normal decomposition of $\mathbf{F}$ (Appendix~\ref{app:norm-decomp-non-normal}). 
Assuming that $\bm{\sigma}_{\min}$ is a stable fixed point and thus $\mathbf{K}$ is positive definite, then $\mathbf{K}_\text{eff}$ is also positive definite, which can be proved in analogy to Appendix~\ref{subsec:pos-semidef_nonzero-trace}.  
Accordingly, we compute the MSD from the energy minimum at steady state as $\bigl\langle \left|\bm{\sigma}-\bm{\sigma}_{\min}\right|^{2}\bigr\rangle 	=T\sum_{n}\lambda_n^{-1}(\mathbf{K}_{\text{eff}})$, where $\{\lambda_n(\mathbf{K}_\text{eff})\}$ are the eigenvalues of $\mathbf{K}_\text{eff}$.
For low values of $T$, which keep the system near the energy minimum at steady state and thus the force approximately linear, our stability analysis recapitulates measurements in simulations after the system has been allowed some time to relax to the target pattern [Fig.~\ref{fig:hopfield}(d)].
The agreement between theory and simulation suggests that nonreciprocity [Eq.~\eqref{eq:antisym}] enhances pattern retrieval by stabilizing the energy minimum associated with the target pattern.
As the number of stored patterns increases, the agreement is quickly lost, presumably due to a concomitant increase in the role of nonlinearity in the interactions.

\section{Intuition for stabilization by nonreciprocal coupling}
\label{sec:intuition}
\begin{figure}
\centering\includegraphics{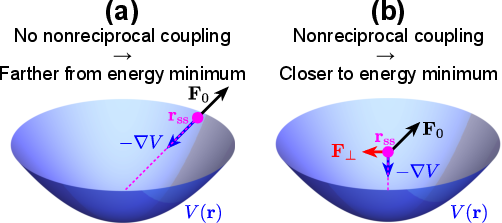}
\caption{%
Schematic illustrating the physical intuition for why nonreciprocal coupling via a transverse force stabilizes active systems whose stable configurations minimize the (reciprocal) interaction energy.
(a) In the absence of nonreciprocal coupling, the system settles at a steady-state position $\mathbf{r}_\text{ss}$ such that the potential force $-\nabla V$ exactly balances the constant force $\mathbf{F}_0$. 
(b) When a transverse force $\mathbf{F}_\perp$ nonreciprocally couples the particles, a smaller $-\nabla V$ is required to counteract $\mathbf{F}_0$.
As a result, $\mathbf{r}_\text{ss}$ lies closer to the energy minimum.
\label{fig:intuition}}
\end{figure}

So far, we have shown that nonreciprocal coupling via a transverse force enhances stability in a diverse set of active systems, whose stable configurations coincide with minima of their (reciprocal) interaction energy.
To provide physical intuition for this effect, we consider a two-dimensional nonreciprocal harmonic oscillator subjected to a constant force $\mathbf{F}_0$, which represents the persistent noise in the limit of infinite persistence time $\tau\rightarrow \infty$.
The dynamics is given by
\begin{equation}
    \dot{\mathbf{r}} = -k\mathbf{r} - \mathbf{A}\mathbf{r} + \mathbf{F}_0,
\end{equation}
where the first term is the gradient $-\nabla V$ of the potential $V = k|\mathbf{r}|^2 /2$, and the second term is the transverse force $\mathbf{F}_\perp$ determined by the antisymmetric force constant matrix
\begin{equation}
    \mathbf{A} = \begin{pmatrix}0 & a\\
-a & 0
\end{pmatrix}.
\end{equation} 
The system converges to a steady-state position $\mathbf{r}_\text{ss} \equiv \mathbf{r}(t\rightarrow\infty)$ satisfying
\begin{equation}
    |\mathbf{r}_\text{ss}|^2=\frac{|\mathbf{F}_0|^2}{k^{2}+a^{2}}.
\end{equation}
As the strength of the nonreciprocal coupling $a$ increases, the steady-state position moves closer to the energy minimum $\mathbf{r} = \mathbf{0}$. 
This behavior can be understood through a simple force-balance argument: a smaller potential force is required to counteract the constant force when the transverse force is present (Fig.~\ref{fig:intuition}).

\section{Nonreciprocal active swimmers}
\label{sec:wca}
%
In general, the stable configurations of active systems need not be energy minima but can arise from the persistent motion of the particles.
To preview the effects of nonreciprocity on active systems with such nonequilibrium stable configurations, we apply a transverse force of the form given by Eq.~\eqref{eq:antisym} to active (i.e., persistently migrating) swimmers with purely repulsive (reciprocal) interactions.
This paradigmatic model of active matter exhibits the classic nonequilibrium phenomenon of motility-induced phase separation (MIPS)~\cite{cates_motility-induced_2015}, in which particles separate into regions of low density and high density due to their persistent motion. 

\begin{figure*}
\centering\includegraphics{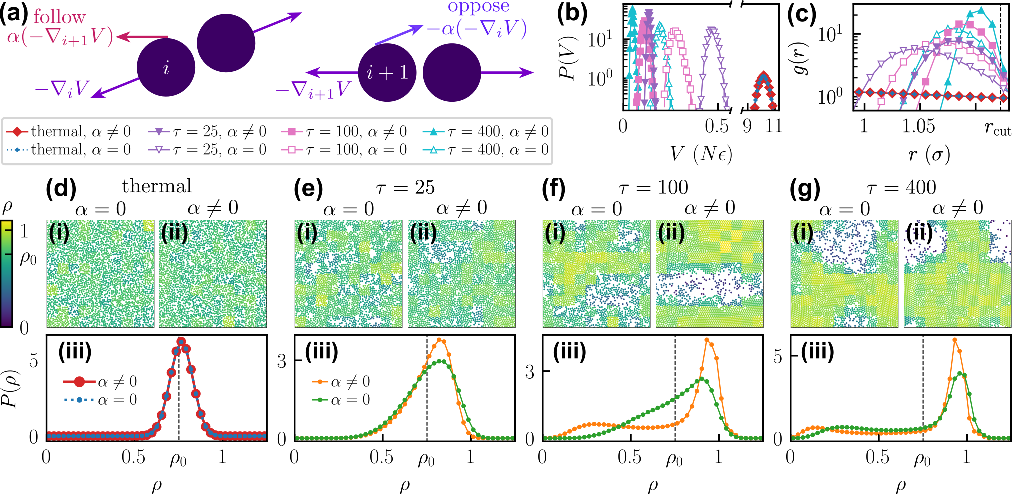}
\caption{Nonreciprocal active swimmers.  
(a) Schematic diagram of the interparticle forces. 
In addition to short-range repulsion (purple arrows) mediated by the WCA potential ($V$), the particles experience nonreciprocal interactions with strength $\alpha$. 
Illustrated is the case of $\alpha > 0$, in which the nonreciprocity drives particle $i$ to follow the motion of particle $i+1$ (blue arrow) and particle $i+1$ to do the opposite of particle $i$ (red arrow). 
(b) Steady-state distribution of potential energy, $P(V)$. 
(c) Steady-state pair correlation function, $g(r)$, zoomed in around the cutoff distance beyond which the pairwise repulsion vanishes ($r_\text{cut}$, dashed line). 
In (b) and (c), results are shown for various noise types and values of the nonreciprocity parameter $\alpha = 0, 4$.
(d-g) Steady-state properties related to density (subpanels i-iii) for persistence times of $\tau = 0$ (thermal noise) (d), $\tau = 25$ (e), $\tau = 100$ (f), and $\tau = 400$ (g).
(i, ii) Representative particle configurations for $\alpha = 0$ (i) and $\alpha = 4$ (ii).
The colors indicate the local density ($\rho$) in each region formed by discretizing the simulation box into a $10 \times 10$ square grid. 
(iii) Distribution of local density, $P(\rho)$.
The total number density ($\rho_0$) is indicated by the dashed line.
In (b), (c), and panels (iii), error bars (vertical solid lines) are shown but, for most of the data, are too small to be seen. 
In (b)-(g), the other parameters are $N=52^2$, $\sigma = 1$, $\epsilon = 0.025$, $L=60$, and $T=1$. 
\label{fig:wca}}
\end{figure*}

We consider $N$ particles interacting (reciprocally) according to the Weeks-Chandler-Andersen (WCA) pairwise potential~\cite{weeks_role_1971} [Fig. \ref{fig:wca}(a)],
\begin{equation}
V=\sum_{i<j} V_\text{WCA}(r_{ij}),
\end{equation}
where $\mathbf{r}=(x_1,y_1,\dots,x_N, y_N)$ denotes the particle positions, and $r_{ij}=|(x_i, y_i) - (x_j, y_j)|$ is the distance between particles $i$ and $j$, and 
\begin{equation}
V_\text{WCA}(r_{ij})=\begin{cases}
4\epsilon\left[\left(\frac{\sigma}{r_{ij}}\right)^{12}-\left(\frac{\sigma}{r_{ij}}\right)^{6}\right], & r_{ij}\leq r_\text{cut}\\
0, & r_{ij}>r_\text{cut}.
\end{cases}
\end{equation}
The particles repel each other within the cutoff distance of $r_\text{cut} \equiv 2^{1/6}\sigma$ but move freely otherwise. 
For sufficiently large persistence time $\tau$, the particles undergo MIPS [Fig.~\ref{fig:wca}, (d)-(g)].

For the nonreciprocal coupling, we apply a transverse force [Eq.~\eqref{eq:antisym}] with antisymmetric matrix 
\begin{equation}
A_{u_i, v_j} = \alpha\delta_{u,v}\left(\delta_{i+1,j}-\delta_{i-1,j}\right)
\label{eq:a-wca}
\end{equation}
for spatial coordinates $u,v\in \{x,y\}$ and particle indices $i,j\in \{1,\dots,N\}$ with periodic boundary conditions $N+1 \equiv 1$ and $0 \equiv N$.
This choice of $\mathbf{F}_\perp$ has strength $\alpha$ and causes each pair of particles $i$ and $i+1$ to nonreciprocally ``follow'' one another.
For example, if $\alpha > 0$ [Fig. \ref{fig:wca}(a)], then particle $i$ tries to move along the direction of the potential force experienced by particle $i+1$, while particle $i+1$ tries to move counter the direction of the potential force experienced by particle $i$. 

At steady state, we observe that the nonreciprocity reduces the total potential energy [Fig.~\ref{fig:wca}(b)] by separating particles which are close enough to repel each other [Fig.~\ref{fig:wca}(c)], consistent with what we have seen for other systems.
Although this effect holds over a wide range of persistence times $\tau$, the resulting density profiles are highly dependent on $\tau$ [Fig.~\ref{fig:wca}, (d)-(g)]. 
At relatively low $\tau$, nonreciprocity breaks up regions of high density, thereby favoring mixing [Fig.~\ref{fig:wca}(e)]. 
At intermediate $\tau$, for which the system is stuck at the onset of MIPS if all interactions are reciprocal, nonreciprocity enables MIPS to go to completion [Fig.~\ref{fig:wca}(f)]. 
When $\tau$ is high enough for MIPS to be stable, nonreciprocity enhances the phase separation, converting some of the remaining regions of intermediate density into regions of low and high density [Fig.~\ref{fig:wca}(g)]. 

Taken together, these results suggest that nonreciprocity might play a stabilizing role for active swimmers, though in a more nuanced sense than in systems whose stable configurations correspond to energy minima. 
Because MIPS emerges only beyond a critical persistence time $\tau$, we may regard the uniform-density and phase-separated states as the relevant stable configurations of the system at low and high $\tau$, respectively.
In both regimes, the nonreciprocal coupling drives the system toward these respective steady states---suppressing clustering when $\tau$ is low and enhancing phase separation when $\tau$ is high---and thus acts as an effective stabilizing mechanism.
Further work will be needed to fully establish this mechanism.

\section{Conclusions}
We have revealed how breaking the reciprocity of interactions by a transverse force shapes the stationary properties of active particles.
The transverse force, which is orthogonal to the gradient of the interaction potential,
helps the system stay at its stable configurations.
This is caused by an interplay between the nonreciprocal couplings and the persistent noise that propels the particles. 
In contrast, under thermal noise the transverse force has no impact on the steady-state distribution of particle positions, indicating that previously reported \cite{bartnick_structural_2015, saha_scalar_2020, chiu_phase_2023,  loos_long-range_2023} changes to the thermal distribution arise from nonreciprocal forces which have finite overlap with the potential gradient. 

We have demonstrated that nonreciprocity can protect a variety of active systems---whose stable configurations minimize their potential energy---against noise, by stiffening springs, aligning spins, and improving associative memory.
We have also presented preliminary evidence that nonreciprocity may play a similar stabilizing role in systems whose stable configurations are genuinely nonequilibrium in origin, in particular, active swimmers exhibiting MIPS at high persistence times. 
Future work should further test this effect, as well as clarify its generality by exploring, for example, attractive reciprocal interactions between swimmers, other systems whose nonlinear interactions can generate nonequilibrium stable configurations (e.g., nonlinear oscillators with a quartic interaction potential~\cite{martin_statistical_2021}), and transverse forces of the form given by Eq.~\eqref{eq:antisym} with different choices of $\mathbf{A}$. 

Broadly, our findings suggest that the consumption of energy by individual particles influences how, if at all, breaking reciprocity in their couplings appears in the stationary states of active matter. 
Although we have focused on AOUPs, we expect this work to be relevant to other models of active particles.
For example, we show in Appendix~\ref{app:abp} that our results on active swimmers (Fig.~\ref{fig:wca}) remain qualitatively unchanged for active Brownian particles~\cite{fily_athermal_2012} (Fig.~\ref{fig:abp}).

For experimental realization, we propose implementing transverse nonreciprocal forces through feedback control~\cite{takatori_feedback_2025}. 
Specifically, the applied field would be dynamically updated using a force-inference scheme that determines the instantaneous interaction forces from measured particle positions. 
The feedback loop must operate on timescales shorter than the characteristic dynamical timescales of the system to ensure that the applied field remains transverse. 
For example, the force-inference model could be learned from experimental data~\cite{yu_physics_2025}, while the transverse force could be implemented optically in Janus particles~\cite{muinos-landin_reinforcement_2021} or biological microswimmers~\cite{arlt_painting_2018,frangipane_dynamic_2018}.

\section*{Acknowledgments}
This work was mainly supported by DOE BES Grant No. DE-SC0019765 by funding to M.D. and S.V. 
A.G. is supported by an EMBO Postdoctoral Fellowship (ALTF 259-2022) and by the National Science Foundation, through the Biophysics of Nuclear Condensates grant (MCB-2044895).
M.D. thanks Yael Avni, Deb Sankar Banerjee, Agnish Behera, Sihan Chen, Corentin Coulais, Efi Efrati, Carlos Floyd, Michel Fruchart, Daiki Goto, Tali Khain, Rituparno Mandal, Qinghao Mao, Billie Meadowcroft, Yuqing Qiu, Gregory Rassolov, Nicolas Romeo, Daniel Seara, Vincenzo Vitelli, and Brady Wu for useful discussions.
This work was completed in part with resources provided by 
the University of Chicago's Research Computing Center.

\section*{Data Availability}
Data or codes supporting the findings of this paper are available from the corresponding author upon reasonable request.

\appendix

\section{Numerical simulations}
\label{app:numerical}
For each system studied in this work, we simulate steady-state properties by numerically integrating Eq.~\eqref{eq:eom}.
Unless otherwise noted, the simulations are run with a time step of $\Delta t = 0.002$ for a total time of $10^8 \Delta t$, where the positions of the particles are stored at every $200\Delta t$ starting from the initial time.
The steady-state properties are obtained as an average over the stored positions.
In the following subsections, we describe the system-specific details.

\subsection{Nonreciprocal harmonic oscillators}
\label{subsec:numerical-harmonic}
For harmonic oscillators (Sec.~\ref{sec:harmonic} and Appendices~\ref{app:robust} and~\ref{app:dusty-plasma}), the system is initialized at $\mathbf{r}=0$.

\subsection{Nonreciprocal spherical models}
\label{subsec:numerical-spherical}
For the spherical models with one species of spins (Appendix~\ref{app:spherical-1}), the spin values are stored only after $500\Delta t$ of equilibration time following initialization of each spin at $\sigma_i = \text{sign}(h)$.
For the spherical model with two species of spins (Sec.~\ref{sec:spherical}), the storage of spin values employs the same equilibration time, and the initial condition is $\bm\sigma = \mathbf{0}$.

\subsection{Nonreciprocal spherical Hopfield model}
\label{subsec:numerical-hopfield}
In simulations of pattern retrieval, we initialize the system such that the overlap $m = \frac{1}{N}\bm{\sigma}\cdot\bm{\xi}^\text{target}$ with the target pattern $\bm{\xi}^\text{target}$ starts approximately at $m_0 = 0.7$; without loss of generality,
we choose the target pattern to be pattern 1, $\bm{\xi}^\text{target} = \bm{\xi}^1$. 
Specifically, we randomly choose the initial state using a modified version of the ``warm" initialization in \cite{mignacco_stochasticity_2021},
\begin{equation}
\bm{\sigma}(0)=m_{0}\frac{\sqrt{N}}{|\bm{\xi}^\text{target}|}\bm{\xi}^\text{target}+r\mathbf{z},
\label{eq:warm}
\end{equation}
where all $z_i \sim \mathcal{N}(0, 1)$ are independent and identically distributed random variables, and $r$ is determined as a solution to $|\bm{\sigma}(0)|^{2}=N$.
Averaging over all possible $\mathbf{z}$, the mean initial overlap with the target pattern would be $m_{0}$ if $|\mathbf{\bm{\xi}}^\text{target}|^2 = N$.
To account for the fact that the target pattern does not actually satisfy this normalization condition while the spin vector (approximately) does (due to the soft spherical constraint; see Sec.~\ref{sec:hopfield}), we have included the factor $\sqrt{N} / |\bm{\xi}^\text{target}|$ in the first term. 
Following initialization, we evolve the system with a time step of $\Delta t = 0.001$ for a total time of $(2\times 10^5)\Delta t$.
After allowing the system to relax to the target pattern in the first half of the simulation, we store the spin values ($\bm\sigma$) at every $100\Delta t$ during the second half.
We use the stored spin values to calculate the steady-state MSD from the energy minimum corresponding to the target pattern (see below).
The MSD is then used to compare simulations with linear stability theory (Sec. \ref{sec:hopfield}).
We also use the spin values at the final time to calculate the final overlap with the target pattern.

To obtain the energy minimum corresponding to a target pattern, we run gradient descent on $V$ [Eq. \eqref{eq:eom} with $\mathbf{F}_\perp = \mathbf{0}$ and $T = 0$] starting from the normalized target pattern, $\bm\sigma(0) = (\sqrt{N}/|\bm\xi^\text{target}|)\bm\xi^\text{target}$.
We run the calculation using a time step of $\Delta t = 0.001$ (same as the pattern retrieval simulations; see above) for a total time of $(1\times 10^5)\Delta t$.

The results shown in Fig. \ref{fig:hopfield} for each set of parameters are an average over 25 pattern sets and 10 initial states per pattern set.
In  Fig. \ref{fig:hopfield}(c), retrieval is deemed possible for a set of parameters if the mean final overlap with the target pattern exceeds 0.5. 

\subsection{Nonreciprocal active swimmers}
\label{subsec:numerical-wca}
The following numerical details are used for simulating active swimmers consisting of AOUPs (Sec.~\ref{sec:wca}) and ABPs (Appendix~\ref{app:abp}).
Simulations are run with a time step $\Delta t$ that depends on the noise persistence time $\tau$, which determines the time scale of the dynamics.
We use $\Delta t = 5 \times 10^{-6}$ for $\tau = 0$ (thermal noise) and $\Delta t = (5 \times 10^{-6})\tau$ for $\tau > 0$.
The system is initialized in a hexagonal lattice that completely fills the $L \times L$ square simulation box (periodic boundaries).
After an equilibration time of $(2 \times 10^{7}) \Delta t$, the system is evolved for another $(2 \times 10^{7}) \Delta t$, during which the particle positions are stored at every $8000\Delta t$ for subsequent use in the calculation of steady-state properties. 
The results shown in Figs.~\ref{fig:wca} and~\ref{fig:abp} reflect data from 5 different simulations for each set of parameters. 

\section{Derivation of $P(\mathbf{r})$ for nonreciprocal harmonic oscillators}
\label{app:prob-harmonic}
In this section, we derive the probability distribution $P(\mathbf{r})$ for a set of harmonic oscillators that move according to Eq.~\eqref{eq:eom} in the main text,
\begin{equation}
\dot{\mathbf{r}} = 
\mathbf{F}(\mathbf{r}) + \bm \eta (t),
\tag{\ref{eq:eom}}
\end{equation}
where the force $\mathbf{F}(\mathbf{r})=-\mathbf{C}\mathbf{r}$ has a non-reciprocal component (i.e., $\mathbf{C}$ is asymmetric) and $\bm \eta (t)$ is persistent noise with zero mean and covariance $\langle \eta_{i} (t) \eta_{j} (t') \rangle = \frac{T}{\tau} \delta_{ij} e^{-|t-t'|/\tau}$.
In Subsection \ref{sec:fpe-pd}, we present the derivation, which yields $P(\mathbf{r}) \propto \exp[-V_\text{eff}(\mathbf{r})/T]$, where $V_\text{eff}(\mathbf{r}) = \frac{1}{2}\mathbf{r}^T \mathbf{K}_\text{eff} \mathbf{r}$ is an effective harmonic potential and $\mathbf{K}_\text{eff}$ is the effective force constant matrix in Eq.~\eqref{eq:k-eff}.
In Subsection \ref{sec:alt_derivation}, we present an alternative derivation of $\mathbf{K}_\text{eff}$ for cases where $\mathbf{C}$ is positive definite (i.e., the system is stable, with all eigenmodes having positive eigenvalue) and normal (i.e., commutes with its transpose). 

\subsection{Derivation of $P(\mathbf{r})$}
\label{sec:fpe-pd}
Here, we derive the probability distribution $P(\mathbf{r})$ using an approach which is almost identical to that in Ref.~\cite{fodor_how_2016}.
As described below, a seemingly small difference in the ansatz for $P(\mathbf{r})$ will be important for determining the correct dependence of $P(\mathbf{r})$ on $\mathbf{F}_\perp$.
To gain some physical insights midway into the derivation, only at the very end do we plug in the explicit forms of the potential, $V=\frac{1}{2}\mathbf{r}^T \mathbf{K} \mathbf{r}$, and transverse force, $\mathbf{F}_\perp=-\mathbf{A}\mathbf{r}$, obtained by the normal decomposition \cite{kwon_structure_2005,noh_steady-state_2015} of $\mathbf{F}$ (see main text).

The derivation begins by recasting the equation of motion \eqref{eq:eom}, which describes overdamped dynamics, as an underdamped Langevin equation. 
To do this, we write down the equation of motion for the noise variable, an Ornstein-Uhlenbeck process given by 
\begin{equation}
\tau\dot{\bm\eta}=-\bm\eta+\bm\zeta(t),
\label{eq:eom-eta}
\end{equation}
where $\bm{\zeta}$ is thermal noise with temperature $T$,
\begin{subequations}
\begin{align}
\langle\zeta_{i}(t)\rangle	&=0,\\
\langle\zeta_{i}(t)\zeta_{j}(t')\rangle	&=2T\delta_{ij}\delta(t-t').
\end{align}
\label{eq:zeta}%
\end{subequations}
Taking the time derivative of Eq. \eqref{eq:eom} and plugging in Eq. \eqref{eq:eom-eta} for $\dot{\bm\eta}$ yields the underdamped Langevin equation,
\begin{equation}
\tau\dot{\mathbf{p}}=\mathbf{F}-\left(\mathbf{I}-\tau\mathbf{J}_{\mathbf{F}}\right)\mathbf{p}+\bm{\zeta}(t),
\label{eq:eom-underdamped}
\end{equation}
where $\mathbf{J}_{\mathbf{F}}$ is the Jacobian of $\mathbf{F}$, with elements $(\mathbf{J}_{\mathbf{F}})_{ij} = \partial F_i / \partial r_j$, $\mathbf{I}$ is the identity matrix, and $\mathbf{p}=\dot{\mathbf{r}}$ are the velocities.

After rescaling time as $\tilde{t}\equiv t/\sqrt{\tau}$ and accordingly the velocities as $\tilde{\mathbf{p}}\equiv \sqrt{\tau}\mathbf{p}$, we convert the Langevin equation to a Fokker-Planck equation~\cite{risken_fokker-planck_1996},
\begin{align}
\dot{P}(\mathbf{r},\tilde{\mathbf{p}})=&\left[-\tilde{\mathbf{p}}^{T}\nabla_{\mathbf{r}}-\mathbf{F}^{T}\nabla_{\tilde{\mathbf{p}}}+\frac{1}{\sqrt{\tau}}\nabla_{\tilde{\mathbf{p}}}\cdot\left(\tilde{\mathbf{p}}-\tau\mathbf{J}_{\mathbf{F}}\tilde{\mathbf{p}}\right)\right. \nonumber\\
&\;\;\left. +\frac{T}{\sqrt{\tau}}\nabla_{\tilde{\mathbf{p}}}^{2}\right]P(\mathbf{r},\tilde{\mathbf{p}}),
\label{eq:fp}
\end{align}
where $P(\mathbf{r},\tilde{\mathbf{p}})$ is the distribution of positions and rescaled velocities.
At steady state, the stationary distribution satisfies Eq.~\eqref{eq:fp} with the left-hand side set to zero.

To solve for the stationary distribution, consider the ansatz
\begin{equation}
P(\mathbf{r},\tilde{\mathbf{p}})\propto\exp\left[-\frac{|\tilde{\mathbf{p}}|^{2}}{2T}-\frac{V}{T}+\sum_{n=1}^{\infty}\tau^{n/2}\psi_{n}(\mathbf{r},\tilde{\mathbf{p}})\right],
\label{eq:ansatz}
\end{equation}
which is similar to that of \cite{fodor_how_2016}, except that the summation starts at $n=1$ instead of $n=2$, and $\tau$ can be arbitrarily large. 
The equation describing the stationary steady-state becomes
\begin{equation}
\sum_{m=0}^{\infty}\tau^{m/2}\phi_{m}[\bm\psi]=0,
\label{eq:fp-ss}
\end{equation}
where $\bm\psi \equiv \{\psi_n (\mathbf{r},\tilde{\mathbf{p}})\}_{n=1}^\infty$ and
\begin{align}
\phi_{0}[\bm\psi] =&\;\frac{1}{T}\tilde{\mathbf{p}}^{T}\mathbf{F}_{\perp}-\tilde{\mathbf{p}}^{T}\nabla_{\tilde{\mathbf{p}}}\psi_{1}+T\nabla_{\tilde{\mathbf{p}}}^{2}\psi_{1},
\\
\phi_{1}[\bm\psi]=&-\tilde{\mathbf{p}}^{T}\nabla_{\mathbf{r}}\psi_{1}
-\mathbf{F}^{T}\nabla_{\tilde{\mathbf{p}}}\psi_{1}
-\text{Tr}\mathbf{J}_{\mathbf{F}}
+\frac{1}{T}\tilde{\mathbf{p}}^{T}\mathbf{J}_{\mathbf{F}}^{T}\tilde{\mathbf{p}}\nonumber\\
&-\tilde{\mathbf{p}}^{T}\nabla_{\tilde{\mathbf{p}}}\psi_{2}
+T\left(\nabla_{\tilde{\mathbf{p}}}\psi_{1}\right)^{T}\nabla_{\tilde{\mathbf{p}}}\psi_{1}
+T\nabla_{\tilde{\mathbf{p}}}^{2}\psi_{2},
\\
\phi_{m}[\bm\psi]=
&-\tilde{\mathbf{p}}^{T}\nabla_{\mathbf{r}}\psi_{m}
-\mathbf{F}^{T}\nabla_{\tilde{\mathbf{p}}}\psi_{m}
-\tilde{\mathbf{p}}^{T}\mathbf{J}_{\mathbf{F}}^{T}\nabla_{\tilde{\mathbf{p}}}\psi_{m-1}
\nonumber\\
&-\tilde{\mathbf{p}}^{T}\nabla_{\tilde{\mathbf{p}}}\psi_{m+1}\nonumber\\
&+T\sum_{n+n'=m+1;\ n,n'\geq1}\left(\nabla_{\tilde{\mathbf{p}}}\psi_{n}\right)^{T}\left(\nabla_{\tilde{\mathbf{p}}}\psi_{n'}\right)\nonumber\\
&+T\nabla_{\tilde{\mathbf{p}}}^{2}\psi_{m+1},\qquad m=2,3,\dots
\end{align}
Using the defining properties of $\mathbf{F}_\perp$ [Eqs. \eqref{eq:perp}-\eqref{eq:div-free}], we recursively solve the equations $\phi_m [\bm\psi] = 0$ for the functions $\bm\psi$, which results in 
\begin{align}
P(\mathbf{r},\tilde{\mathbf{p}}) &\propto\exp\Bigg(-\frac{\left|\tilde{\mathbf{p}}-\sqrt{\tau}\mathbf{F}_{\perp}\right|^{2}}{2T}-\frac{V}{T}-\frac{\tau}{2T}\tilde{\mathbf{p}}^{T}\mathbf{H}_{V}\tilde{\mathbf{p}}\Bigg.\nonumber\\
&\qquad\qquad\Bigg.-\frac{\tau}{2T}|\nabla_{\mathbf{r}}V|^{2}\Bigg)
\label{eq:p-rp}
\end{align}
as the steady-state distribution of positions and rescaled velocities. 
Our result, which is exact for harmonic forces (i.e., linear in $\mathbf{r}$), can be seen as an approximation for generic forces if the persistence time $\tau$ is small.
Indeed, if $\mathbf{F}$ is nonlinear, the corrections will be $O(\tau^{n/2})$ where $n \geq 3$, though solving for them is highly nontrivial in general, even if one chooses a different ansatz for $P(\mathbf{r}, \tilde{\mathbf{p}})$~\cite{bonilla_active_2019, martin_statistical_2021}. 
Importantly, we would like to point out that $\mathbf{F}_\perp$ appears at $O(\sqrt{\tau})$ (to lowest order in $\tau$), a finding that would not have been obtainable had the summation in Eq.~\eqref{eq:ansatz} started instead at $n=2$ as in Ref.~\cite{fodor_how_2016}. 
In particular, $\mathbf{F}_\perp$ shifts the velocities [Eq.~\eqref{eq:p-rp}, first term in parentheses] and thus acts analogously to a (magnetic) vector potential.

Since the velocities are also coupled by $V$ [Eq. \eqref{eq:p-rp}, third term in parentheses], the influence of $\mathbf{F}_\perp$ persists in the marginal distribution $P(\mathbf{r})$.
Indeed, integrating $P(\mathbf{r},\mathbf{p})$ [Eq. \eqref{eq:p-rp}] over all $\tilde{\mathbf{p}}$ and using the integral identity (Ref.~\cite{zee_quantum_2010}, p. 15, Eq. 22)
\begin{align}
\int_{-\infty}^{\infty}d\mathbf{x}\exp\left(-\frac{1}{2}\mathbf{x}^{T}\mathbf{M}\mathbf{x}+\mathbf{v}^{T}\mathbf{x}\right)=&\;\sqrt{\frac{(2\pi)^{N}}{\det\mathbf{M}}}\nonumber\\
&\times\exp\left(\frac{1}{2}\mathbf{v}^{T}\mathbf{M}^{-1}\mathbf{v}\right)
\end{align}
for matrix $\mathbf{M}\in\mathbb{R}^{N\times N}$ and vector $\mathbf{v}\in\mathbb{R}^{N}$, we arrive at
\begin{align}
    P(\mathbf{r}) = &\;\frac{1}{Z} \left[\det\left(\mathbf{I}+\tau\mathbf{H}_V\right)\right]^{-\frac{1}{2}} \nonumber\\
    &\times\exp\bigg[ - \frac{V}{T} - \frac{\tau}{2T}|\nabla_{\mathbf{r}}V|^2 \bigg.\nonumber\\
    &\qquad\qquad\!\!\!\bigg.+ \frac{\tau}{2T} \mathbf{F}_\perp^T \left((\mathbf{I} + \tau \mathbf{H}_V)^{-1} - \mathbf{I}\right) \mathbf{F}_\perp\bigg] \, ,
\label{eq:p-lin-det}
\end{align}
which can also be written as
\begin{align}
    P(\mathbf{r}) = &\;\frac{1}{Z} \exp\bigg[ - \frac{V}{T} - \frac{\tau}{2T}|\nabla_{\mathbf{r}}V|^2 \nonumber\\
    &\qquad\qquad\!\!\bigg.+ \frac{\tau}{2T} \mathbf{F}_\perp^T  \left((\mathbf{I}+\tau \mathbf{H}_V)^{-1} - \mathbf{I}\right) \mathbf{F}_\perp \bigg.\nonumber\\
    &\qquad\qquad\!\!\bigg. -\frac{1}{2}\tr\ln\left(\mathbf{I}+\tau\mathbf{H}_V\right) \bigg] \, .
\label{eq:p-lin}
\end{align}
Here, we have absorbed the constant $(2\pi T)^{N/2}$ from the Gaussian integration into the factor $Z$ which normalizes the probability distribution.
For sufficiently small $\tau$, one can expand
\begin{align}
    P(\mathbf{r}) \approx &\;\frac{1}{Z} \exp\bigg( 
    - \frac{V}{T} 
    - \frac{\tau}{2T}|\nabla_{\mathbf{r}}V|^2 
    - \frac{\tau}{2}\nabla^2 V \bigg.\nonumber\\
    &\qquad\qquad\!\!\bigg.- \frac{\tau^2}{2T} \mathbf{F}_\perp^T  \mathbf{H}_V  \mathbf{F}_\perp 
    \bigg) \, .
\label{eq:p-expand}
\end{align}
%
We note that for general $\mathbf{F}$, there is a correction to Eqs.~\eqref{eq:p-lin-det}-\eqref{eq:p-expand} at $\mathcal{O}(\tau^2)$, which is characterized by the anharmonic part of the potential, i.e., the part with a nonvanishing contribution to $\nabla_\mathbf{r}^3 V$.

To arrive at a more explicit solution for $P(\mathbf{r})$, we substitute $V = \frac{1}{2}\mathbf{r}^T \mathbf{K} \mathbf{r}$ and $\mathbf{F}_\perp = -\mathbf{A}\mathbf{r}$, as obtained by normal decomposition \cite{kwon_structure_2005,noh_steady-state_2015} of $\mathbf{F}$ (see main text), into Eq.~\eqref{eq:p-lin}.
This yields $P(\mathbf{r}) \propto \exp(-V_\text{eff}/T)$, where $V_\text{eff}(\mathbf{r}) = \frac{1}{2}\mathbf{r}^T \mathbf{K}_\text{eff} \mathbf{r}$ is an effective harmonic potential and $\mathbf{K}_\text{eff}$ is the effective force constant matrix in Eq.~\eqref{eq:k-eff}.

For one choice of non-reciprocal forces presented in the main text,
\begin{equation}
\mathbf{F}_\perp = \mathbf{A}(-\nabla_\mathbf{r} V),
\tag{\ref{eq:antisym}}
\end{equation}
the last term in the square brackets in Eq.~\eqref{eq:p-expand} can be written as $\frac{\tau^2}{2T} \mathbf{F}_\perp^T  \mathbf{H}_V  \mathbf{F}_\perp = \frac{\tau^2}{2T} (\nabla_\mathbf{r} V)^T \mathbf{A}^T \mathbf{H}_V  \mathbf{A} (\nabla_\mathbf{r} V)$.
Thus, non-reciprocity has an effect similar to increasing the persistence time of the noise by a configuration-dependent $\Delta \tau$, with
\begin{equation}
    \tau\min\lambda(\mathbf{A}^T \mathbf{H}_V  \mathbf{A}) \leq \frac{\Delta \tau}{\tau} \leq \tau\frac{\tr(\mathbf{A}^T \mathbf{H}_V  \mathbf{A})}{N} \, ,
\end{equation}
where $\lambda$ refers to the set of eigenvalues of its argument.

\subsection{Alternative derivation of $\mathbf{K}_\text{eff}$}
\label{sec:alt_derivation}
Next, we present an alternative derivation of the effective force constant matrix $\mathbf{K}_\text{eff}$ [Eq. \eqref{eq:k-eff}] for cases where the force constant matrix $\mathbf{C}$ is  positive definite (i.e., the system is stable, with all eigenmodes having positive eigenvalue) and normal (i.e., commutes with its transpose). 
We note that the first part of the derivation holds for all positive definite $\mathbf{C}$. 
Only in the second part of the derivation do we additionally assume that $\mathbf{C}$ is normal.

Since $\mathbf{F}=-\mathbf{C}\mathbf{r}$ is linear in $\mathbf{r}$, we can use standard tools from linear stochastic calculus~\cite{gardiner_handbook_1985} to analytically solve Eq.~\eqref{eq:eom} for any driving force $\bm \eta (t)$:
\begin{equation}
    \mathbf{r}(t) = \int_{-\infty}^{t} \! dt' \, e^{-\mathbf{C} (t-t')} \bm \eta (t') \, ,
\end{equation}
where we have assumed that the last explicitly known configuration lies infinitely far in the past.
Thus, the average displacement from the rest configuration vanishes, $\langle \mathbf{r}(t) \rangle = 0$, and the covariance matrix describing the fluctuations around the rest configuration is
\begin{align}
    \left\langle \mathbf{r}(t)  \mathbf{r}^T(t) \right\rangle = \int_{-\infty}^{t} \! dt' \, \int_{-\infty}^{t} \! dt'' \, &\Big[e^{-\mathbf{C} (t-t')} \left\langle \bm \eta (t') \bm \eta^T (t'') \right\rangle\Big.\nonumber\\
    &\ \Big.\times e^{-\mathbf{C}^T (t-t'')}\Big] \, .
\end{align}
Substituting the exponentially correlated noise, and changing the integration variables, leads to
\begin{equation}
    \left\langle \mathbf{r}(t) \mathbf{r}^T(t) \right\rangle = \frac{T}{\tau} \int_{0}^{\infty} \! dt' \, \int_{0}^{\infty} \! dt'' \, e^{-\mathbf{C} t'} e^{-\mathbf{C}^T t''} \,  e^{-|t' - t''|/\tau} \, .
\end{equation}
Splitting the integral into the two cases $t'>t''$ and $t'\leq t''$, followed by another change of variables, leads to
\begin{align}
    \left\langle \mathbf{r}(t) \mathbf{r}^T(t) \right\rangle = 
    &\;\frac{T}{\tau} \int_{0}^{\infty} \! dt' \, e^{(-\mathbf{C} - \mathbf{I}/\tau) t'}
    \, \int_{0}^{\infty} \! dt'' \, e^{-\mathbf{C} t''} e^{-\mathbf{C}^T t''} \nonumber\\ 
    &+
    \left(\frac{T}{\tau} \int_{0}^{\infty} \! dt'' \, e^{-\mathbf{C} t''} e^{-\mathbf{C}^T t''} \right. \nonumber\\
    &\qquad\!\left.\times\int_{0}^{\infty} \! dt' \,  e^{(-\mathbf{C}^T-\mathbf{I}/\tau)t'}\right) \, .
\end{align}
Assuming that $\mathbf{C}$ is positive definite, one can carry out two of the integrals to find:
\begin{align}
    \left\langle \mathbf{r}(t)\mathbf{r}^T(t) \right\rangle = 
    &\;T (\mathbf{I}+\tau\mathbf{C})^{-1}
    \int_{0}^{\infty} \! dt'' \, e^{-\mathbf{C} t''} e^{-\mathbf{C}^T t''} \nonumber\\
    &+
    T \int_{0}^{\infty} \! dt'' \, e^{-\mathbf{C} t''} e^{-\mathbf{C}^T t''} (\mathbf{I}+\tau\mathbf{C}^T)^{-1} \, .
\end{align}

So far, the only assumption we have made about $\mathbf{C}$ is that it is positive semidefinite. 
Hereafter, we assume that $\mathbf{C}$ is a normal matrix and thus commutes with its transpose.
This leads to 
\begin{align}
    \left\langle \mathbf{r}(t)\mathbf{r}^T(t) \right\rangle = 
    &\;T \left[ (\mathbf{I}+\tau\mathbf{C})^{-1}
    (\mathbf{C}+\mathbf{C}^T)^{-1} \right.\nonumber\\
    &\quad\ \ \left.+
     (\mathbf{C}+\mathbf{C}^T)^{-1} (\mathbf{I}+\tau\mathbf{C}^T)^{-1} \right] \, .
\end{align}
Now, we will determine the effective stiffness matrix of this system, by using 
\begin{equation}
    \left\langle \mathbf{r}(t) \mathbf{r}^T(t) \right\rangle \coloneqq T \mathbf{K}_\text{eff}^{-1} \, .
\end{equation}
Using the decomposition $\mathbf{C} = \mathbf{K} + \mathbf{A}$ where $\mathbf{K}$ is symmetric and $\mathbf{A}$ is anti-symmetric, and exploiting that $\mathbf{C}$ commutes with its transpose and that $\mathbf{K}$ commutes with $\mathbf{A}$, one has
\begin{equation}
    \mathbf{K}_\text{eff} = 
    \mathbf{K} 
    (\mathbf{I}+\tau\mathbf{K})
    +
    \tau\mathbf{A}^T
    \left[\mathbf{I} +(\tau\mathbf{K})^{-1} \right]^{-1}     \mathbf{A}
    \, .
\end{equation}
This is equivalent to $\mathbf{K}_\text{eff}$ in Eq.~\eqref{eq:k-eff}.

\section{Proof: the third term of $\mathbf{K}_\text{eff}$ further stabilizes the potential energy minima}\label{app:third-term-stabilizes}
In this section, we prove that the third term of  $\mathbf{K}_\text{eff}$ [Eq. \eqref{eq:k-eff}], which corresponds to the transverse force $\mathbf{F}_\perp = -\mathbf{A}\mathbf{r}$ that couples the particles and is hereafter referred to as $\mathbf{K}_\text{eff}^{(3)}$, further stabilizes the minima of the potential $V$.
As discussed in the main text, the effective force constant matrix $\mathbf{K}_\text{eff}$ governs $P(\mathbf{r})$ under persistent noise through the effective potential $V_\text{eff}(\mathbf{r})= \frac{1}{2}\mathbf{r}^T \mathbf{K}_\text{eff}\mathbf{r}$, while $V(\mathbf{r}) = \frac{1}{2}\mathbf{r}^T \mathbf{K}\mathbf{r}$ governs $P(\mathbf{r})$ under thermal noise.
We assume that $\mathbf{K}$ is positive semidefinite, implying that the system is not unstable.
Also, we suppose that $\mathbf{K}\mathbf{A}\neq \mathbf{0}$, which should be true in general and is used to prove that $\mathbf{K}_\text{eff}^{(3)}$ is nonzero.

In Secs. \ref{subsec:pos-semidef_nonzero-trace}-\ref{subsec:zero-modes-same}, we prove intermediate results.
In Sec. \ref{subsec:finish-proof}, we use these results to finally prove the main result. 

\subsection{Proof: the third term is positive semidefinite and has nonzero trace}\label{subsec:pos-semidef_nonzero-trace}
To prove that $\mathbf{K}_\text{eff}^{(3)}$ is positive semidefinite and nonzero, it is useful to write it as [cf. Eq. \eqref{eq:k-eff}] 
\begin{equation}
\mathbf{K}_\text{eff}^{(3)} = \mathbf{A}^T \mathbf{S} \mathbf{A},
\label{eq:kasa}
\end{equation}
where we have defined
\begin{equation}
\mathbf{S}\equiv \tau\left[\mathbf{I}-\left(\mathbf{I}+\tau\mathbf{K}\right)^{-1}\right].
\label{eq:s}
\end{equation}

We first prove that $\mathbf{K}_\text{eff}^{(3)}$ is positive semidefinite.
We make multiple uses of the fact that  
\begin{align}
&\text{matrix }\mathbf{M}
\text{ is positive semidefinite} \nonumber \\
&\Leftrightarrow \quad
\mathbf{M}=\mathbf{M}_{1/2}^T \mathbf{M}_{1/2}
\text{ for some matrix }
\mathbf{M}_{1/2}.
\label{eq:fact-psd-sq}
\end{align}
Given the set $\{\lambda\}$ of eigenvalues of $\mathbf{K}$, the set of eigenvalues of $\mathbf{S}$ is $\left\{\tau\left[1-(1+\tau\lambda)^{-1}\right]\right\}$ [Eq. \eqref{eq:s}]. 
So, $\mathbf{S}$ is positive semidefinite by the positive semidefiniteness of $\mathbf{K}$.
Using the forward direction of fact \eqref{eq:fact-psd-sq}, we can write $\mathbf{S}=\mathbf{S}_{1/2}^T \mathbf{S}_{1/2}$ for some matrix $\mathbf{S}_{1/2}$.
Therefore, by the backward direction of fact \eqref{eq:fact-psd-sq}, $\mathbf{K}_\text{eff}^{(3)}=(\mathbf{S}_{1/2}\mathbf{A})^T (\mathbf{S}_{1/2}\mathbf{A})$ is also positive semidefinite.

Next, we prove that our assumption of  $\mathbf{K}\mathbf{A} \neq \mathbf{0}$ implies  $\mathbf{K}_\text{eff}^{(3)}\neq \mathbf{0}$.
Proceeding by contradiction, we suppose that $\mathbf{K}_\text{eff}^{(3)}=\mathbf{0}$. 
By Eq. \eqref{eq:kasa} and the positive definiteness of $\mathbf{S}$, we have $\mathbf{S}\mathbf{A} = \mathbf{0}$ [\cite{abadir_matrix_2005}, p. 221, Exercise 8.27(c)].
Plugging in the definition of $\mathbf{S}$ [Eq. \eqref{eq:s}] and rearranging, we find $\mathbf{A}=\left(\mathbf{I}+\tau\mathbf{K}\right)^{-1}\mathbf{A}$.
Multiplying both sides by $\mathbf{I}+\tau \mathbf{K}$ and rearranging, we obtain $\mathbf{K}\mathbf{A}=\mathbf{0}$, a contradiction.
Hence,  $\mathbf{K}_\text{eff}^{(3)}\neq \mathbf{0}$.

Since $\mathbf{K}_\text{eff}^{(3)}$ is both positive semidefinite and nonzero, then $\text{tr}\,\mathbf{K}_\text{eff}^{(3)}\neq0$  [\cite{abadir_matrix_2005}, p. 214, Exercise 8.8(a)].

\subsection{Proof: the third term increases at least one eigenvalue of $\mathbf{K}_\text{eff}$ while keeping the others the same}\label{subsec:eval-geq}
Having proved that $\mathbf{K}_\text{eff}^{(3)}$ is positive semidefinite and has nonzero trace (Sec. \ref{subsec:pos-semidef_nonzero-trace}), we are ready to elucidate the effect of this term on the eigenvalues of $\mathbf{K}_\text{eff}$. 
Throughout this subsection, when we refer to the $n$th eigenvalue of some matrix, the index $n$ denotes the position within an ascending (or descending) ordering of the eigenvalues.

Using the fact that the $n$th eigenvalue of the sum of two positive semidefinite matrices is greater than or equal to the $n$th eigenvalue of either matrix (\cite{abadir_matrix_2005}, p. 346, Exercise 12.45), we find that the $n$th eigenvalue of $\mathbf{K}_{\text{eff}}$ either stays the same or increases when adding $\mathbf{K}_\text{eff}^{(3)}$ to the rest of the matrix. 
Combining this result with the fact that $\mathbf{K}_\text{eff}^{(3)}$ has nonzero trace, it follows that this term increases at least one eigenvalue (for some $n$) of the effective force constant matrix.
Thus, $\mathbf{K}_\text{eff}^{(3)}$ increases at least one eigenvalue of $\mathbf{K}_\text{eff}$ while keeping the remaining eigenvalues unchanged.

\subsection{Proof: the third term does not change the zero modes of $\mathbf{K}_\text{eff}$}\label{subsec:zero-modes-same}
We now prove that $\mathbf{K}_\text{eff}^{(3)}$ does not change the zero modes of $\mathbf{K}_\text{eff}$.
As shown in \cite{kwon_structure_2005}, we can express $\mathbf{A}=\mathbf{Q}\mathbf{K}$. 
Plugging this into Eq. \eqref{eq:k-eff}, we can readily see that the zero modes of $\mathbf{K}$ are also zero modes of $\mathbf{K}_\text{eff}^{(3)}$.
Thus, the zero modes of $\mathbf{K}_\text{eff}$ remain as zero modes after adding $\mathbf{K}_\text{eff}^{(3)}$, since the zero modes of the other terms in $\mathbf{K}_\text{eff}$ [Eq. \eqref{eq:k-eff}] are exactly those of $\mathbf{K}$. 
Since $\mathbf{K}_\text{eff}^{(3)}$ does not decrease the eigenvalues of $\mathbf{K}_\text{eff}$ (Sec. \ref{subsec:eval-geq}), then $\mathbf{K}_\text{eff}^{(3)}$ also does not cause $\mathbf{K}_\text{eff}$ to have additional zero modes. 
Therefore, $\mathbf{K}_\text{eff}^{(3)}$ has no effect on the zero modes of $\mathbf{K}_\text{eff}$, which are exactly those of $\mathbf{K}$. 

\subsection{Proof: the third term further stabilizes the potential energy minima}\label{subsec:finish-proof}
We now complete our proof that $\mathbf{K}_\text{eff}^{(3)}$ further stabilizes the minima of $V(\mathbf{r}) = \frac{1}{2}\mathbf{r}^T \mathbf{K}\mathbf{r}$.
First, notice that these minima, which satisfy $V(\mathbf{r})=0$, include not only $\mathbf{r}=0$ but also all $\mathbf{r}\neq 0$ that are linear combinations of the zero modes of $\mathbf{K}$. 
From Sec. \ref{subsec:zero-modes-same}, we have that $\mathbf{K}_\text{eff}^{(3)}$ does not change the zero modes of $\mathbf{K}_\text{eff}$.
Combining this result with that of Sec. \ref{subsec:eval-geq}, we find that $\mathbf{K}_\text{eff}^{(3)}$ additionally increases at least one of the nonzero eigenvalues of $\mathbf{K}_\text{eff}$ while leaving the remaining nonzero eigenvalues unchanged.
Therefore, $\mathbf{K}_\text{eff}^{(3)}$ makes the system more likely to be found at the minima of $V$.  

\section{Average potential under persistent noise and normal decomposition}
\label{app:avg-pot}
In this section, we study the evolution of the quasi-potential $V(\mathbf{r})$, averaged over fluctuations, for particles moving according to Eq.~\eqref{eq:eom} with force $\mathbf{F}$ admitting the normal decomposition of Eqs.~\eqref{eq:f-norm-decomp}-\eqref{eq:div-free}.
The dynamics of the average potential is given by
\begin{equation}
    \frac{d}{dt}\langle V(\mathbf{r}) \rangle 
    = \langle \nabla V(\mathbf{r}) \cdot \dot{\mathbf{r}} \rangle
    = \langle \nabla V(\mathbf{r}) \cdot \mathbf{F}(\mathbf{r}) \rangle
    + \langle \nabla V(\mathbf{r}) \cdot \bm\eta (t) \rangle \,.
\end{equation}
After substituting Eq.~\eqref{eq:f-norm-decomp} and using Eq.~\eqref{eq:perp}, one has
\begin{equation}
    \frac{d}{dt}\langle V(\mathbf{r}) \rangle 
    = -\langle |\nabla V|^2 \rangle
    + \langle \nabla V(\mathbf{r}) \cdot \bm\eta (t) \rangle \,.
\end{equation}
To evaluate the last term, we exploit the fact that the noise $\bm\eta$ is Gaussian and apply the Novikov-Furutsu theorem~\cite{novikov_1965}:
\begin{widetext}
\begin{align}
    \frac{d}{dt}\langle V(\mathbf{r}) \rangle 
    & = -\langle |\nabla V|^2 \rangle
    + \sum_{i,n} \int_{-\infty}^{t} dt' \left\langle \frac{\delta}{\delta \eta_n(t')}\frac{\partial}{\partial r_i} V(\mathbf{r}) \right\rangle \left\langle\eta_n(t') \eta_i (t) \right\rangle \\
    & = -\langle |\nabla V|^2 \rangle
    + \sum_{i,n,m} \int_{-\infty}^{t} dt' \left\langle \frac{\partial^2 V(\mathbf{r})}{\partial r_i \partial r_m} \, \frac{\delta r_m(t)}{\delta \eta_n(t')} \right\rangle \left\langle\eta_n(t') \eta_i (t) \right\rangle \,.
\label{eq:potential_time_derivative}
\end{align}
\end{widetext}
We now apply linear response theory to calculate $\delta r_m / \delta \eta_n$.
First, we consider an expansion of Eq.~\eqref{eq:eom} close to some point $\mathbf{r}_0$ that represents the state of the system at time $t_0$.
We can then approximate the force, $\mathbf{F}(\mathbf{r}) \approx \mathbf{J}_{\mathbf{F}} \cdot (\mathbf{r} - \mathbf{r}_0)$, and write the dynamics of deviations from this point as
\begin{equation}
\delta\dot{\mathbf{r}} \approx 
\mathbf{J}_{\mathbf{F}} \cdot \delta\mathbf{r} + \bm \eta (t),
\label{eq:perturbation}
\end{equation}
where we have defined $\delta\mathbf{r} \coloneqq \mathbf{r} - \mathbf{r}_0$ and $\delta\mathbf{r}(t_0) = 0$.
Solving this equation using linear stochastic calculus~\cite{gardiner_handbook_1985}, we obtain
\begin{equation}
    \delta\mathbf{r}(t) = \int_{t_0}^{t} \! dt' \, e^{\mathbf{J}_{\mathbf{F}} (t-t')} \cdot \bm \eta (t') \, ,
    \label{eq:lin-resp}
\end{equation}
where $\mathbf{J}_{\mathbf{F}}$ is the Jacobian of $\mathbf{F}$, with elements $(\mathbf{J}_{\mathbf{F}})_{ij} \coloneqq \partial F_i / \partial r_j$.
If we consider $\mathbf{r}[\bm\eta]$ as a functional over the time domain and use the linear response result of Eq. \eqref{eq:lin-resp}, we have
\begin{equation}
    \frac{\delta r_m(t)}{\delta \eta_n(t')} = \left(e^{\mathbf{J}_{\mathbf{F}} (t-t')}\right)_{mn} \, ,
    \label{eq:lin-resp-variation}
\end{equation}
which is valid in the vicinity of $\mathbf{r}_0$ and remains useful as long as nonlinearities are small.
It is natural to take $\mathbf{r}_0$ as a minimum of the potential landscape $V(\mathbf{r})$.
Substituting into Eq.~\eqref{eq:potential_time_derivative}, together with the covariance of the noise, $\langle \eta_{i} (t) \eta_{n} (t') \rangle = \frac{T}{\tau} \delta_{in} e^{-|t-t'|/\tau}$, gives
\begin{align}
    \frac{d}{dt}\langle V(\mathbf{r}) \rangle 
    = &-\langle |\nabla V|^2 \rangle\nonumber\\
    &+ \frac{T}{\tau} \sum_{nm} \left\langle\frac{\partial^2 V(\mathbf{r})}{\partial r_n \partial r_m} \, \left[\int_{0}^{\infty} dz e^{(\mathbf{J}_{\mathbf{F}} - \mathbf{I}/\tau)z}\right]_{mn} \right\rangle \,,
    \label{eq:potential_time_derivative_sub}
\end{align}
where we have used the coordinate transform $t-t' = z$.
Note that for large $t-t' = z$, which are certainly part of the integration domain, Eq.~\eqref{eq:lin-resp-variation} becomes inaccurate.
Nonetheless, Eq.~\eqref{eq:potential_time_derivative_sub} remains useful since the weight of these contributions exponentially approaches zero as $e^{-z/\tau}$.
After carrying out the integral, we have
\begin{align}
    \frac{d}{dt}\langle V(\mathbf{r}) \rangle 
    = &-\langle |\nabla V|^2 \rangle\nonumber\\
    &+ T \sum_{nm} \left\langle \frac{\partial^2 V(\mathbf{r})}{\partial r_n \partial r_m} \, \left[(\mathbf{I} -\tau\mathbf{J}_{\mathbf{F}})^{-1}\right]_{mn} \right\rangle \,.
\end{align}
We define the Hessian of the potential as $\mathbf{H}_V$, with elements $(\mathbf{H}_V)_{ij} \coloneqq \partial^2 V / (\partial r_i \partial r_j)$.
Thus, we finally arrive at
\begin{equation}
    \frac{d}{dt}\langle V(\mathbf{r}) \rangle 
    = -\langle |\nabla V|^2 \rangle
    + T \, \left\langle \mathbf{H}_V \colon  (\mathbf{I} -\tau\mathbf{J}_{\mathbf{F}})^{-1} \right\rangle\,,
\end{equation}
where $\colon$ is the operator for the Frobenius inner product.
Note that the first term on the right-hand side is strictly negative.
Hence, the deterministic part of the dynamics will gradually reduce the average potential as long as a normal decomposition exists.
The second term on the right-hand side, $T \, \left\langle \mathbf{H}_V \colon  (\mathbf{I} -\tau\mathbf{J}_{\mathbf{F}})^{-1} \right\rangle = T \left\langle \tr[\mathbf{H}_V \cdot  (\mathbf{I} -\tau\mathbf{J}_{\mathbf{F}})^{-1}] \right\rangle$, is strictly positive which can be seen as follows.
First, we define $\mathbf{B} \coloneqq (\mathbf{I} -\tau\mathbf{J}_{\mathbf{F}})^{-1}$.
Theorem~1 in Ref.~\cite{fang_1994} states that $\min\lambda(\overline{\mathbf{B}}) \, \tr\mathbf{H}_V \leq \tr[\mathbf{B} \cdot \mathbf{H}_V]\leq \max\lambda(\overline{\mathbf{B}}) \, \tr\mathbf{H}_V$ where $\lambda(\overline{\mathbf{B}})$ refers to the set of eigenvalues of $\overline{\mathbf{B}} \coloneqq (\mathbf{B} + \mathbf{B}^T)/2$.
Near the point $\mathbf{r}_0$ about which we have linearized $\mathbf{F}(\mathbf{r})$ [Eq.~\eqref{eq:lin-resp}], if we assume the system is stable, the symmetric matrix $\mathbf{H}_V$ is positive semidefinite, and hence $\tr\mathbf{H}_V > 0$.
Next, we will analyze
\begin{align}
    \overline{\mathbf{B}} 
    &= \frac{1}{2} \left[\mathbf{B} + \mathbf{B}^T\right] 
    = \frac{1}{2} \left[(\mathbf{I} -\tau\mathbf{J}_{\mathbf{F}})^{-1} + (\mathbf{I} -\tau\mathbf{J}^T_{\mathbf{F}})^{-1}\right]\nonumber\\
    &=  (\mathbf{I} -\tau\mathbf{J}_{\mathbf{F}})^{-1} \cdot \left[ \mathbf{I} -\tau \overline{\mathbf{J}}_{\mathbf{F}}\right] \cdot (\mathbf{I} -\tau\mathbf{J}_{\mathbf{F}})^{-1,T} \, .
\end{align}
Because $-\overline{\mathbf{J}}_\mathbf{F} \coloneqq -(\mathbf{J}_\mathbf{F} + \mathbf{J}^T_\mathbf{F})/2$ is also positive semidefinite for a stable system, and $\mathbf{Q} \cdot \mathbf{M} \cdot \mathbf{Q}^T$ is positive semidefinite for arbitrary matrices $\mathbf{Q}$ as long as $\mathbf{M}$ is positive semidefinite\footnote{To immediately see this, consider $\mathbf{x}^T \cdot \mathbf{Q} \cdot \mathbf{M} \cdot \mathbf{Q}^T \cdot \mathbf{x} = \mathbf{y}^T \cdot \mathbf{M} \cdot \mathbf{y}$ where $\mathbf{y} \coloneqq \mathbf{Q}^T \cdot \mathbf{x}$.}, it follows that $\overline{\mathbf{B}}$ has only positive eigenvalues.
Taken together, this means that $\tr[\mathbf{B} \cdot \mathbf{H}_V] \geq 0$.
Hence, fluctuations inject energy into the system.

To understand the effect of persistence and nonreciprocity on the energy-injecting term $T \left\langle \tr[\mathbf{H}_V \cdot  (\mathbf{I} -\tau\mathbf{J}_{\mathbf{F}})^{-1}] \right\rangle$, we hereafter assume that $-\mathbf{J}_{\mathbf{F}}$ is a normal matrix, implying a normal decomposition 
$-\mathbf{J}_{\mathbf{F}} \coloneqq \mathbf{K} + \mathbf{A}$ (see main text) where $\mathbf{K}=-\overline{\mathbf{J}}_\mathbf{F}$ and $\mathbf{A} = -(\mathbf{J}_\mathbf{F} - \mathbf{J}^T_\mathbf{F})/2$ are the symmetric and antisymmetric parts of $-\mathbf{J}_{\mathbf{F}}$, respectively~\cite{noh_steady-state_2015}.
We also linearize the potential part of the interaction force, $-\nabla V$, as we have done for the total force $\mathbf{F}$ [Eq.~\eqref{eq:lin-resp}], leading to $\mathbf{H}_V = \mathbf{K}$ and thus
\begin{equation}
    \left\langle \tr[\mathbf{H}_V \cdot  (\mathbf{I} - \tau\mathbf{J}_\mathbf{F})^{-1}] \right\rangle = \tr[\mathbf{K} \cdot  (\mathbf{I} + \tau\mathbf{K} + \tau\mathbf{A})^{-1}].
\end{equation}
Since $-\mathbf{J}_{\mathbf{F}}$ is a real-valued normal matrix, the following hold:
\begin{enumerate}
    \item $-\mathbf{J}_{\mathbf{F}}$ is diagonalizable [Ref.~\cite{horn_matrix_2012}, Theorem 2.5.3(b)]; 
    \item $-\mathbf{J}_{\mathbf{F}}$ can have real and complex (i.e., nonzero imaginary part) eigenvalues, where the latter come in pairs of complex conjugates $\{\lambda, \lambda^*\}$ (Ref.~\cite{horn_matrix_2012}, Theorem 2.5.8);
    \item the $n$th eigenvector of $-\mathbf{J}_{\mathbf{F}}$, with corresponding eigenvalue $\lambda_n$, is also an eigenvector of $\mathbf{K}$ and $\mathbf{A}$ with eigenvalues $\lambda_n (\mathbf{K}) = \text{Re}\lambda_n$ and $\lambda_n (\mathbf{A}) = i\text{Im}\lambda_n$, respectively (as follows straightforwardly from Ref.~\cite{grone_normal_1987}, Conditions 31 and 33-36).
\end{enumerate}
Using these properties, we obtain
\begin{widetext}
\begin{equation}
    \tr[\mathbf{K} \cdot  (\mathbf{I} + \tau\mathbf{K} + \tau\mathbf{A})^{-1}]
    = \sum_{n} \lambda_n(\mathbf{K}) \frac{1 + \tau\lambda_n(\mathbf{K})}{[1+\tau\lambda_n(\mathbf{K})]^2 + |\tau\lambda_n(\mathbf{A})|^2},
    \label{eq:tr-eigen}
\end{equation}
\end{widetext}
implying 
\begin{equation}
    \tr[\mathbf{K} \cdot  (\mathbf{I} + \tau\mathbf{K} + \tau\mathbf{A})^{-1}]
    < \tr[\mathbf{K} \cdot  (\mathbf{I} + \tau\mathbf{K})^{-1}] 
    < \tr\mathbf{K}.
\end{equation}
With increasing $\tau$, the term $\tr[\mathbf{K} \cdot  (\mathbf{I} + \tau\mathbf{K} + \tau\mathbf{A})^{-1}]$ becomes smaller and less energy is injected.
If $\tau \neq 0$, then non-reciprocity ($\mathbf{A}$) also reduces the energy injected into the system. 

\section{Robustness to disorder and open boundary conditions}
\label{app:robust}
\begin{figure*}
\centering\includegraphics{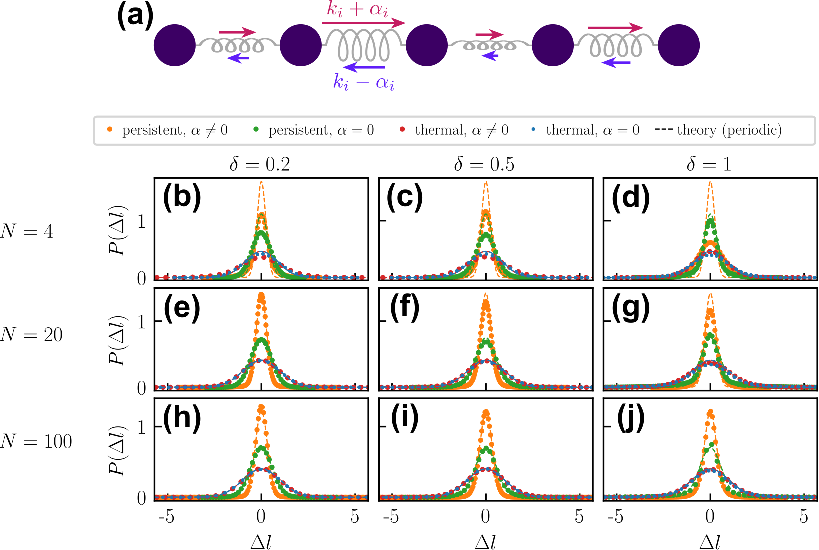}
\caption{Nonreciprocal spring-mass chain with disorder and open boundary conditions. 
(a) Schematic diagram.
(b-j) Steady-state distribution [$P(\Delta l)$] of displacements of the springs from their rest length ($\Delta l$)  for various noise types and values of nonreciprocity parameter ($\alpha=0,2$). 
These results are shown for chain lengths (b-d) $N=4$, (e-g) $N=20$, and (h-j) $N=100$ and disorder strengths (b, e, h) $\delta=0.2$, (c, f, i) $\delta = 0.5$, and (d, g, h) $\delta=1$.
Each set of simulated values (points of a given color) is calculated from a different disorder realization and overlaid by the theoretical values (dashed line of the same color) for the corresponding periodic system ($\delta = 0$).
The other parameters are $k = 1$, $T=1$, and $\tau = 2$. 
\label{fig:aperiodic-harmonic}}
\end{figure*}

To explore the effect of disorder and open boundary conditions on the nonreciprocal spring-mass chain, we consider the force [Fig. \ref{fig:aperiodic-harmonic}(a)]
\begin{widetext}
\begin{equation}
F_i = \begin{cases}
-(k_{1}-\alpha_{1})(r_{i}-r_{i+1}), & i=1,\\
-(k_{i}-\alpha_{i})(r_{i}-r_{i+1})-(k_{i-1}+\alpha_{i-1})(r_{i}-r_{i-1}), & i=2,\dots,N-1,\\
-(k_{N-1}+\alpha_{N-1})(r_{N}-r_{N-1}) & i=N,
\end{cases}
\label{eq:f-lin-1-obc}
\end{equation}
\end{widetext}
where 
\begin{align}
k_i &\sim U\big((1-\delta)k, (1+\delta)k\big) \\
\alpha_i &\sim U\big((1-\delta)\alpha, (1+\delta)\alpha\big)
\end{align}
are random variables drawn from a uniform distribution on the intervals $[(1-\delta)k, (1+\delta)k]$ and $[(1-\delta)\alpha, (1+\delta)\alpha]$.
Fig. \ref{fig:aperiodic-harmonic}(b)-(j) shows that as the system size ($N$) increases, the steady-state distribution of spring displacements becomes less affected by disorder and, in fact, converges to the corresponding distribution of the periodic analog ($\delta = 0$).

\section{Numerical normal decomposition for nonreciprocal harmonic oscillators with non-normal force constant matrix}
\label{app:norm-decomp-non-normal}
In this section, we describe how we numerically implement the method of Ref.~\cite{kwon_structure_2005}, Sec. `Decomposition of the Force', to perform the normal decomposition of $\mathbf{F}=-\mathbf{C}\mathbf{r}$ when the force constant matrix $\mathbf{C}$ is asymmetric (i.e., nonreciprocal interactions) and not normal (i.e., does not commute with its transpose). 
As we discussed in the main text, the normal decomposition uniquely yields~\cite{ao_stochastic_2003, kwon_structure_2005, noh_steady-state_2015} $\mathbf{F}=-\nabla V + \mathbf{F}_\perp$,
where $V = \frac{1}{2}\mathbf{r}^T \mathbf{K} \mathbf{r}$ for symmetric matrix $\mathbf{K}$ and $\mathbf{F}_\perp = -\mathbf{A}\mathbf{r}$ for traceless matrix $\mathbf{A}$ such that $\mathbf{K}\mathbf{A}$ is antisymmetric.
If $\mathbf{C}$ is a normal matrix, then $\mathbf{K}$ and $\mathbf{A}$ are simply the symmetric and antisymmetric parts of $\mathbf{C}$, respectively~\cite{noh_steady-state_2015}.
However, if $\mathbf{C}$ is not normal, this result does not hold, and we instead numerically calculate $\mathbf{K}$ and $\mathbf{A}$ using the aforementioned method (\cite{kwon_structure_2005}, Sec. `Decomposition of the Force'), which we implement as follows:
\begin{enumerate}
\item Compute the eigenvalues $\{\lambda_{n}\}$ and normalized right eigenvectors $\{\mathbf{v}_{n}\}$ of $\mathbf{C}$ using standard numerical diagonalization. 
\label{item:diag}
\item Compute the normalized left eigenvectors $\{\mathbf{u}_{n}^{T}\}$ according to $\mathbf{U}^T = \mathbf{V}^{-1}$, where $\mathbf{U}^T$ is the matrix whose rows are $\{\mathbf{u}_{n}^{T}\}$ and $\mathbf{V}$ is the matrix whose columns are $\{\mathbf{v}_{n}\}$. (Note: this inverse relation is guaranteed by the diagonalizability of $\mathbf{C}$.)
\item For each $m$ and $n$, solve
\begin{equation}
(\lambda_{m}+\lambda_{n})\tilde{Q}_{mn}=(\lambda_{m}-\lambda_{n})\mathbf{u}_{m}^{T}\mathbf{u}_{n}
\label{eq:q-tilde-uu}
\end{equation}
for $\tilde{Q}_{mn}$.
If $\lambda_m = \lambda_n = 0$, there are infinitely many solutions for $\tilde{Q}_{mn}$, and so we simply set $\tilde{Q}_{mn} = 0$, which satisfies the requirement that $\tilde{\mathbf{Q}}$ is antisymmetric~\cite{kwon_structure_2005}. (See below for other cases where $\lambda_m + \lambda_n = 0$.) 
\label{item:q-tilde}
\item Compute $\mathbf{Q}=\mathbf{V}\tilde{\mathbf{Q}}\mathbf{V}^T$.
\item Compute $\mathbf{K}=(\mathbf{I}+\mathbf{Q})^{-1}\mathbf{C}$ and $
\mathbf{A} = \mathbf{C}-\mathbf{K}$.
\end{enumerate}
We note that this procedure does not work for certain choices of $\mathbf{C}$. 
If $\mathbf{C}$ is not diagonalizable, then Step~\ref{item:diag} cannot be carried out, and one can instead follow the alternative approach of Ref.~\cite{kwon_structure_2005},  Sec. `Appendix', which involves representing $\mathbf{C}$ in Jordan normal form (see~\cite{shi_computation_2023} for a numerical implementation). 
Also, in Step~\ref{item:q-tilde}, if two eigenvalues of $\mathbf{C}$ sum to zero but the right-hand side of Eq.~\eqref{eq:q-tilde-uu} does not vanish, then the corresponding matrix element $\tilde{Q}_{mn}$ cannot be solved for; see Ref.~\cite{kwon_structure_2005}, Sec.~`Singularities', for further discussion.

\section{Spring-mass model of dusty plasma}
\label{app:dusty-plasma}
\begin{figure}
\centering\includegraphics{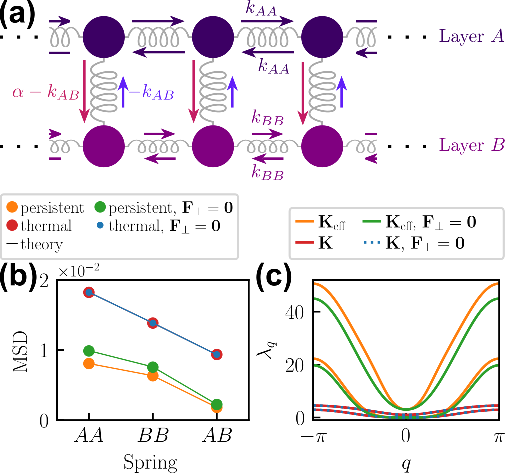}
\caption{Spring-mass model of dusty plasma. 
(a) Schematic diagram. 
Within layer $A$ ($B$), masses are reciprocally coupled to their nearest neighbors via a spring with force constant $k_{AA}$ ($k_{BB}$).
In contrast, each mass of layer $A$ behaves as if it is connected to the corresponding mass of layer $B$ by an unstable spring with force constant $-k_{AB} <0$, while the latter mass is effectively connected to the former mass via a stable spring with force constant $\alpha-k_{AB}>0$.
(b) Mean squared displacement (MSD) of the springs connecting masses of layer $A$ ($AA$), layer $B$ ($BB$), and different layers ($AB$) for various noise types and with or without the transverse force ($\mathbf{F}_\perp$).
Each set of simulated values (points of a given color) is overlaid by the corresponding theoretical values (line of the same color).
(c) Eigenvalues ($\lambda_q$), indexed by wavenumber ($q$) and further sorted into two bands, of the (effective) force constant matrix governing $P(\mathbf{r})$ for various noise types [persistent ($\mathbf{K}_\text{eff}$), thermal ($\mathbf{K}$)] and with or without $\mathbf{F}_\perp$.
In (b)-(c), the parameters are $N=100$, $k_{AA} = 1$, $k_{BB} = 0.6$, $k_{AB}= 1$, $\alpha=3$,  $T=0.01$, and $\tau = 2$. 
\label{fig:dusty-plasma}}
\end{figure}

In this section, we confirm with simulations that the analytical solution $P(\mathbf{r}) \propto \exp[-V_\text{eff}(\mathbf{r}) / T]$, where $V_\text{eff}(\mathbf{r}) = \frac{1}{2}\mathbf{r}^T {K}_\text{eff} \mathbf{r}$ and $\mathbf{K}_\text{eff}$ is given by Eq.~\eqref{eq:k-eff}, is exact also for nonreciprocal harmonic oscillators with non-normal force constant matrix.
In the main text, we numerically validated the solution for a nonreciprocal spring-mass chain [Eq. \eqref{eq:f-lin-1} and Fig. \ref{fig:harmonic}(a)], where the force constant matrix $\mathbf{C}$ (see Sec.~\ref{sec:harmonic}) is normal. 
Here, we consider an experimentally relevant harmonic description of dusty plasma \cite{melzer_structure_1996} [Fig. \ref{fig:dusty-plasma}(a)].
The system consists of two spring-mass chains, which represent two vertically stacked layers of dust particles.
Mass $i=1,\dots,N$ of each layer $l=A,B$ feels a force
\begin{subequations}
\begin{align}
F_{A,i}=&-k_{AA}\sum_{s=\pm 1}(r_{A,i}-r_{A,i+s})-(-k_{AB})(r_{A,i}-r_{B,i}),\\
F_{B,i}=&-k_{BB}\sum_{s=\pm 1}(r_{B,i}-r_{B,i+s})\nonumber\\
&-(\alpha-k_{AB})(r_{B,i}-r_{A,i}),
\end{align}
\label{eq:f-lin-2}%
\end{subequations}
respectively, where $r_{l,i}$ is the displacement of mass $i$ of layer $l$ from its rest position.
Writing $\mathbf{F}=-\mathbf{C}\mathbf{r}$, one can show that the force constant matrix $\mathbf{C}$ is not a normal matrix.
Besides $\mathbf{C}$ not being normal, this system is different from the spring-mass chain of Eq. \eqref{eq:f-lin-1} because there are two species of masses, and reciprocity is broken with a non-transverse force, i.e.,
the conservative part of $\mathbf{F}$ that results from normal decomposition ($-\nabla V$) is also affected by the nonreciprocity parameter $\alpha$.
The last difference becomes evident, e.g., if we set $\alpha = 0$ [Fig. \ref{fig:dusty-plasma}(a)], upon which the interaction between layers turns purely repulsive and thus the system turns unstable. 

Fig. \ref{fig:dusty-plasma}(b) shows the mean squared displacement of the springs from simulations and theory.
The simulations are carried out as described in Sec. \ref{app:numerical}. 
The theoretical calculations involve the normal decomposition of $\mathbf{F}$ [Eq. \eqref{eq:f-lin-2}], which we carry out numerically since $\mathbf{C}$ is not a normal matrix (see \cite{kwon_structure_2005} and Appendix~\ref{app:norm-decomp-non-normal}).
This numerical procedure yields the matrices $\mathbf{K}$ and $\mathbf{A}$ governing the potential $V$ and transverse force $\mathbf{F}_\perp$, respectively (see main text).
For purposes of comparison, we also run simulations setting $\mathbf{A}$ and thus $\mathbf{F}_\perp$ to zero.
The simulations show that the transverse force does not affect the distribution of spring displacements under thermal noise but stabilizes the springs at their rest lengths under persistent noise.
The simulated distributions are in perfect agreement with theory, where the average displacement of each spring from its rest length is determined to be zero, and the MSD of each type of spring is calculated from the numerically computed force constant matrix $\mathbf{K}$ for thermal noise and $\mathbf{K}_\text{eff}$ [Eq. \eqref{eq:k-eff}] for persistent noise.

Further validation of the theory can be obtained by analyzing the eigenvalues and eigenvectors of $\mathbf{K}$ and $\mathbf{K}_\text{eff}$.
Due to the translational symmetry of the system and the coupling between layers [Eq. \eqref{eq:f-lin-2} and Fig. \ref{fig:dusty-plasma}(a)], the normal modes are a superposition of the Fourier normal modes of layers $A$ and $B$ with the same wavenumber $q$. 
Fig.\ref{fig:dusty-plasma}(c) plots the eigenvalues versus $q$. 
As the theory predicts, the stabilizing nature of the transverse force is reflected in how the eigenvalues of the (effective) force constant matrix increase in general, and stay the same otherwise, as we switch from thermal ($\mathbf{K}$) to persistent noise ($\mathbf{K}_\text{eff}$) [Fig. \ref{fig:dusty-plasma}(c)].

\section{Nonreciprocal spherical models with one spin species}
\label{app:spherical-1}
\subsection{Ferromagnetic coupling}\label{subsec:ferromagnetic}
\begin{figure*}
\centering\includegraphics{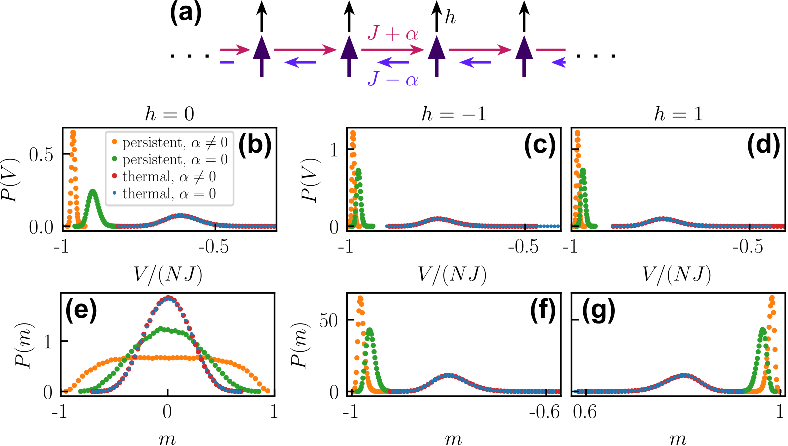}
\caption{Nonreciprocal spherical model. (a) Schematic diagram.
(b-g) Steady-state distributions of potential energy [$P(V)$] (b-d) and magnetization [$P(m)$] for various types of noise (persistent and thermal) and values of the nonreciprocity parameter ($\alpha=0,2$). 
Results are shown for magnetic field strengths (b-e) $h=0$, (c-f) $h=-1$, and (d-g) $h=1$. 
The other parameters are $N=100$, $J = 1$, $\mu = 2$, $T=1$, and $\tau = 2$. 
\label{fig:spherical}}
\end{figure*}

We investigate a spin-based analog~\cite{seara_non-reciprocal_2023} of the nonreciprocal spring-mass chain~\eqref{eq:f-lin-1}.
Specifically, we consider a nonreciprocal version of the spherical model~\cite{berlin_spherical_1952} [Fig.~\ref{fig:spherical}(a)].
Each spin experiences a force 
\begin{align}
F_{i} 
&= -\mu \left(|\bm\sigma|^2- N\right)\sigma_i
+\sum_{s=\pm 1}(J-s\alpha)\sigma_{i+s} + h,
\label{eq:f-sph-1}
\end{align}
where $\bm \sigma$ is the spin vector, $N$ is the number of spins, and periodic boundary conditions are imposed as $\sigma_{N+1} \equiv \sigma_1$ and $\sigma_0 \equiv \sigma_N$.
Spin $i+1$ experiences a force from spin $i$ with coupling constant $J+\alpha$, while spin $i$ feels a force from spin $i+1$ with coupling constant $J-\alpha$.
The magnetic field has strength $h$.
In this model, the nonlinearity [first term of Eq.~\eqref{eq:f-sph-1}] is a soft constraint that imposes the condition $|\bm\sigma|^2 = N$, where $\mu > 0$ is the hardness. 

Fig.~\ref{fig:spherical}, (b)-(g), shows simulations with ferromagnetic interaction ($J>0$). 
As expected by the normal decomposition of $\mathbf{F}$ [Eq. \eqref{eq:f-norm-decomp}-\eqref{eq:div-free}], 
\begin{align}
    V &= \frac{1}{4}\mu \left(|\bm\sigma|^2- N\right)^2 -\sum_i J\sigma_i\sigma_{i+1} - h\sum_i \sigma_i, \\
    F_{\perp, i} &= \alpha(\sigma_{i+1} - \sigma_{i-1}),
\end{align}
nonreciprocity ($\alpha\neq 0$) does not change the steady-state distributions of potential energy $V$ [Fig.~\ref{fig:spherical}, (b)-(d)] and magnetization [Fig.~\ref{fig:spherical}, (e)-(g)] under thermal noise. 
In contrast, under persistent noise, nonreciprocity reduces the potential energy [Fig.~\ref{fig:spherical}, (b)-(d)] while enhancing the alignment of the spins with each other and, for $h\neq 0$, also the magnetic field  [Fig.~\ref{fig:spherical}, (e)-(g)].
Therefore, persistent noise enables nonreciprocity to stabilize the fully aligned states, which globally minimize the potential energy.

\begin{figure}
\centering\includegraphics{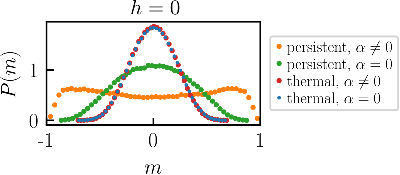}
\caption{Same as Fig.~\ref{fig:spherical}(e) except with $\tau = 4$.
\label{fig:spherical.higher-tau_h0}}
\end{figure}

We would like to comment briefly on the steady-state distribution of magnetization being peaked at 0 for all simulations in Fig.~\ref{fig:spherical}(e), where no magnetic field is present. 
For the reciprocal system under thermal noise, it is known that spontaneous magnetization cannot occur in the thermodynamic limit ($N \rightarrow \infty$)~\cite{berlin_spherical_1952}. 
We believe that this does not explain our observed magnetization distributions, which are obtained using finite-size simulations.
Instead, we attribute our findings to the specific parameters chosen.
For example, increasing the persistence time $\tau$ of the noise causes the nonreciprocal system to have a distribution peaked at nonzero magnetization (Fig.~\ref{fig:spherical.higher-tau_h0}).

\subsection{Antiferromagnetic coupling}
\begin{figure}
\centering\includegraphics{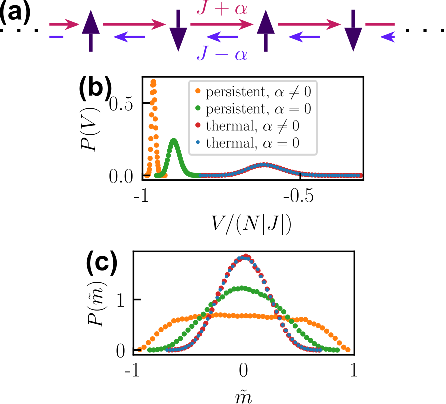}
\caption{Nonreciprocal spherical model with antiferromagnetic coupling. (a) Schematic diagram.
(b-c) Steady-state distributions of potential energy [$P(V)$] (b) and antimagnetization [$P(\tilde{m})$] (c) for various types of noise (persistent and thermal) and values of the nonreciprocity parameter ($\alpha=0,2$).
The other parameters are $N=100$, $J = -1$, $\mu = 2$, $T=1$, and $\tau = 2$. 
\label{fig:afm-spherical}}
\end{figure}

We now explore the nonreciprocal spherical model with antiferromagnetic coupling $J<0$ [Fig.~\ref{fig:afm-spherical}(a)].
Simulations of various steady-state distributions [Fig.~\ref{fig:afm-spherical}, (b)-(c)] show results that are analogous to the ferromagnetic case [Fig.~\ref{fig:spherical}, (b) and (e)]. 
In particular, we observe that nonreciprocity promotes the lowering of the potential energy $V$ [Fig.~\ref{fig:afm-spherical}(b)] and the increase of the antimagnetization [Fig.~\ref{fig:afm-spherical}(c)] under persistent noise.
Hence, even with antiferromagnetic coupling, we still observe the stabilization of the global energy minima (fully antialigned spins) due to the interplay of nonreciprocity and persistent noise.

\section{Active Brownian particles}
\label{app:abp}
\begin{figure*}
\centering\includegraphics{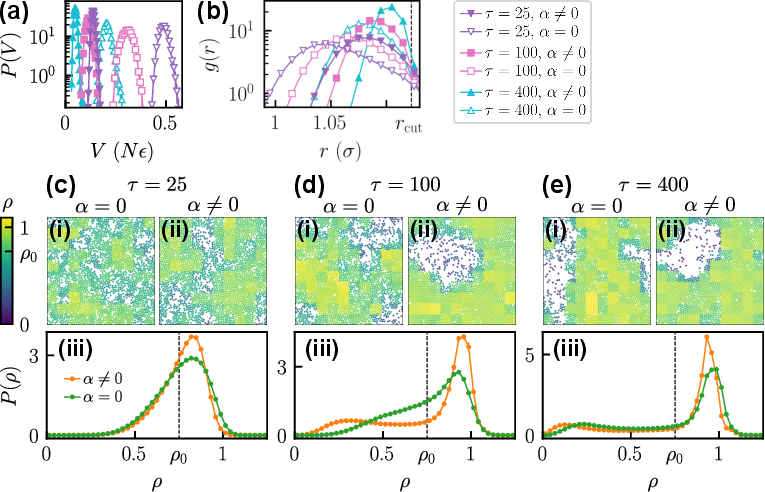}
\caption{Nonreciprocal active swimmers in the active Brownian particle (ABP) model.  
(a) Steady-state distribution of potential energy, $P(V)$. 
(b) Steady-state pair correlation function, $g(r)$, zoomed in around the cutoff distance beyond which the pairwise repulsion vanishes ($r_\text{cut}$, dashed line). 
In (a) and (b), results are shown for various noise types and values of the nonreciprocity parameter $\alpha$.
(c-e) Steady-state properties related to density (subpanels i-iii) for persistence times of $\tau = 25$ (c), $\tau = 100$ (d), and $\tau = 400$ (e).
(i, ii) Representative particle configurations for $\alpha = 0$ (i) and $\alpha = 4$ (ii).
The colors indicate the local density ($\rho$) in each region formed by discretizing the simulation box into a $10 \times 10$ square grid. 
(iii) Distribution of local density, $P(\rho)$.
The total number density ($\rho_0$) is indicated by the dashed line.
In (a), (b), and panels (iii), error bars (vertical solid lines) are shown but, for most of the data, are too small to be seen. 
The other parameters are the same as Fig.~\ref{fig:wca}.
\label{fig:abp}}
\end{figure*}

To show that our results are not limited to AOUPs, we consider nonreciprocal active swimmers within another prototypical model of active matter: active Brownian particles (ABPs)~\cite{fily_athermal_2012}.
In analogy to the AOUP equation of motion \eqref{eq:eom}, the position $\mathbf{r}_i = (x_i, y_i)$ of ABP $i$ evolves as \cite{fodor_how_2016, basu_active_2018}
\begin{align}
\dot{\mathbf{r}}_i &= \mathbf{F}_i(\mathbf{r}) + v_0(\cos\theta_i, \sin\theta_i) \\
\dot{\theta}_i &= \zeta_i(t),
\end{align}
where $\theta_i$ is the orientation of the particle and $v_0$ is the self-propulsion velocity (assumed to be the same for all particles).
The orientation experiences thermal noise $\zeta_i$, which has zero mean and no temporal correlations, $\langle \zeta_{i} (t) \zeta_{j} (t') \rangle = \frac{2}{\tau} \delta_{ij} \delta(t-t')$, where $\tau$ is the persistence time of the propulsion. 
We simulate ABPs using the same procedure as AOUPs and the same value for parameters that are common to both models (see Appendix.~\ref{subsec:numerical-wca}).
For $v_0$, the only parameter of ABPs that is not shared by AOUPs, we set $v_0 = \sqrt{2T/\tau}$, which we obtain by equating the steady-state effective diffusivity $D_\text{eff}=T$ \cite{koumakis_directed_2014, maggi_generalized_2014} of a free ($\mathbf{F}_i=\mathbf{0}$) AOUP to the corresponding $D_\text{eff}=v_0^2\tau / 2$ \cite{basu_active_2018} of a free ABP.
The simulation results of ABPs (Fig.~\ref{fig:abp}) show no qualitative differences to AOUPs [Fig.~\ref{fig:wca}, (b)-(g)].

\bibliographystyle{apsrev4-2}
%

\end{document}